# Coexistence of Diamagnetism and Vanishingly Small Electrical Resistance at Ambient Temperature and Pressure in Nanostructures

Dev Kumar Thapa[1], Saurav Islam[2], Subham Kumar Saha[1], Phanibhusan Singha Mahapatra[2], Biswajit Bhattacharyya[1], T. Phanindra Sai[2], Rekha Mahadevu[1], Satish Patil[1], Arindam Ghosh[2,3] and Anshu Pandey[1]*

**Affiliations:**

[1]Solid State and Structural Chemistry Unit, Indian Institute of Science, Bangalore 560012, India

[2]Department of Physics, Indian Institute of Science, Bangalore 560012, India

[3]Center for Nano Science and Engineering, Indian Institute of Science, Bangalore 560012, India

*Correspondence to: anshup@iisc.ac.in

**The great practical utility has motivated extensive efforts to discover ultra-low resistance electrical conductors and superconductors in ambience. Here we report the observation of vanishingly small electrical resistance at the ambient temperature and pressure conditions in films and pellets of a nanostructured material that is composed of silver particles embedded into a gold matrix. Upon cooling below a sample-specific temperature scale ($T_c$) as high as 286 K, the film resistance drops below ∼ 2 μΩ, being limited by measurement uncertainty. The corresponding resistivity (∼ $10^{-12}$ Ω.m) is at least four orders of magnitude below that of elemental noble metals, such as gold, silver or copper. Furthermore, the samples become strongly diamagnetic below $T_c$, with volume susceptibilities as low as −0.056. We additionally describe methods to tune $T_c$ to temperatures much higher than room temperature.**

Suppressing the scattering of electrons is the key to the absence of resistance (R) in a conductor, which may be achieved via non-trivial topological protection [1,2] or macroscopic coherence. The latter is observed in a superconductor, which has diversified from simple elements such as mercury [3,4], to more recently, materials like cuprates [5,6], iron oxypnictides [7,8], bismuth [9,10], graphene [11] and even $H_2S$ [12]. Despite the discovery of a large number of materials that undergo normal to superconducting transitions, it is apparent that conditions of extremely low temperature (T) and/or extremely high pressure are necessary in each case [12-21]. Therefore, there remains an unfulfilled need for a material system that undergoes this transition under more conveniently attainable T and pressure conditions.

Nanostructured materials have been extensively investigated in the context of superconductivity [22-31] as well as novel many-body phase coherent effects due to spatial confinement [32,33]. For example, recent studies have shown that nanostructuring gives rise to enhancements in both critical magnetic fields [28,29,34] and transition temperatures [35], over their bulk counterparts.

The explanations of transition temperature rise were based on extensions of the BCS formalism [36] as well as more unconventional pictures such as polarization waves and plasmons [25,37-42] that may trigger non-phonon based electron pairing mechanisms (*viz.* plasmonic) [43-46]. Metallic nanostructures have also been shown to host persistent non-dissipative current [47,48], arising from phase coherent circulation of electrons along closed loops, that can impact the magnetic response of these structures in external magnetic fields [49] .

Here we investigated the properties of nanostructures (NS) prepared from Au and Ag. Both materials have low electron-phonon coupling and are not known to exhibit a superconducting state independently. During our studies we synthesized NS comprising of silver particles (~1 nm) embedded into a gold matrix. Sample preparation was done using standard colloidal techniques. Briefly, the method employed by us involves the preparation of silver particles (exemplified in Figure S1 in the Supplementary Information (SI)) and their subsequent incorporation into a gold matrix. Figure 1a and 1b show a representative transmission electron microscopy (TEM) image and a high resolution TEM (HRTEM) image of the resultant particles. The lattice planes observed in Figure 1b correspond to the [111] plane of Au and Ag. Due to the near identical lattice constants of both constituents, lattice diffraction methods lead to a single diffraction maximum for each material. Figure S2a (SI) additionally shows the ensemble X-ray diffraction pattern of the material. The ensemble level reflections are also in excellent agreement to the standard powder patterns of either Au or Ag. To better understand their internal structure, we studied these NS using electron microscopy. Samples were deposited on a carbon coated copper grid for TEM. Figure 1c shows a high annular aperture dark field (HAADF)-elemental contrast image of these NS. The elemental occurrences of silver and gold along the red line are shown in Figure 1d. Figures S2b-e (SI) further exemplify the elemental occurrences within these NS. Figure S3 (SI) additionally shows the overall compositional analysis of the material using energy dispersive x-ray spectroscopy (EDAX). Collectively, these data confirm the successful inclusion of silver nanoparticles into the gold matrix.

For electrical characterization, these NS were cast into films with thickness ranging from 25 nm to of order hundred nm. Figure 2a shows a micrograph of a typical NS film cast on Cr/Au (5 nm/50 nm) leads (on a glass slide). The van der Pauw-like lead geometry allows *R* measurements in multiple orientations of two and four-probe configurations. The details of the film casting procedure are available in Section S1 (Material and Methods) of SI. We ensured appropriate sintering of particles, which is highlighted in Figure S4 for robust electrical connectivity across the film. Following preparation, the films were often encapsulated with various passivating agents for protection against environmental contamination or exposure, especially against oxygen adsorption/oxidation. The details of over 125 devices measured during this work can be found in Table S1 in SI. We observed that in ten of these films, *R* drops below the measurement uncertainty at a characteristic temperature scale. While the unsuccessful results are attributed to oxygen exposure of the samples during preparation and transfer, we found that ensuring the quality of the inert environment, which involves suitable encapsulation of samples prepared in an inert environment with < 20 ppm oxygen, yields more than 50 % successful devices in terms of exhibiting the drop in $R$. Figure 2b shows the $T$-dependence of $R$ in two such NS films P20519FEE_20 and P20519FEE_06 (Table S1, SI), close to their respective transitions. The transition temperatures $T_c$ of 178.8 K (P20519FEE_20) and 272.8 K (P20319FEE_06) are widely different between the two devices, and found to be sensitive to multiple factors including the oxygen exposure, optimal silver nanocluster density, ageing, and inter-NS-grain connectivity (Section S8 - S9, SI). The width of the resistive transition was found to range over ~ 0.2 – 4 K, depending also on aging, environmental exposure and thermal/electrical stress history (Section S11, SI). Figure 2c illustrates this with the broadening (and shifting of $T_c \approx$ 286 K) of the transition in P20319FEE_06 six days after preparation (See Section S19, SI, for more discussions on the nature of $T$-dependence of $R$). The low resistance state was observed to be stable from a few hours to several days (Section S12 in SI), and usually stopped working due to environmental contamination or contact failure.

Given the nanostructured nature of the film, we confirmed that the absence of voltage drop below the transition is not due to physical detachment of the current path from the part of the film coupled to the voltage leads [50]. We observed two-probe $R \sim 10 - 100$ Ω between all lead pairs, as well as a metal-like linear current ($I$)-voltage ($V$) characteristics both above and below the transition (Section S10, SI), which eliminates the possibility of percolative decoupling of the leads. The four-probe $I - V$ characteristics is linear for $T > T_c$ (Figure 2d), but the voltage drop becomes immeasurably small at $T < T_c$ even at the maximum bias current of $\pm 10$ mA. The measurement uncertainty from the distribution of the voltage drop (inset of Figure 2e), sets the minimum measurable $R$ of ~ 2 μΩ in our experiment, indicating that the transition at $T_c$ involves nearly six orders of magnitude decrease in $R$, which is similar to resistive transitions in conventional superconductors (Figure 2e). Importantly, the observed $R$ in the low resistance state corresponds to upper limit of resistivity $\approx 10^{-12}$ Ω.m, which is at least four orders of magnitude lower than that of elemental noble metals (*i.e.* Au, Ag, or Cu) at similar $T$ range.

To investigate the impact of an external magnetic field ($B$) on the transition, we measured the $T$-dependence of $R$ up to 3 Tesla. As shown in Figure 3a for film P11017FE0_02 (Table S1, SI), the transition can be observed even at a field scale of 3 Tesla, although the $T_c$ decreases by about ~ 6 K from its zero field magnitude of 236 K (Figure 3b). In addition, the width of the transition is progressively broadened with increasing $B$, which is a known effect in high-$T_c$ superconducting films [51], where it indicates increase in thermally activated flux flow [52].

Besides the transition in $R$, we observed appearance of strong diamagnetism when the NS are cooled below the $T_c$. To demonstrate this with direct magnetic measurements, we prepared bulk pellets (roughly 6 μL volume) of the NS, and measured the $T$-dependence of the susceptibility in a SQUID magnetometer. Figure 3c shows the volume susceptibility of pellet P11117PE0_01 (Table S1, SI), where a sharp decrease in susceptibility at 238 K can be readily observed in the zero field cooled (ZFC) condition. From the mass (65 mg) and density (11.61 g/cc) of this sample we find diamagnetic susceptibilities as low as -0.056 (SI units) that are about a tenth of what we observe in the case of a 90 mg pellet of lead at 5 K, using the same measurement protocols (Section S20, SI). It is possible that imperfect sintering and the continued persistence of the nanoparticles within the material [53] may have some detrimental effect on the true diamagnetic susceptibility in this class of materials. Regardless, the observed diamagnetism is far stronger than the values associated with most normal materials, as well as with previous reports of nanostructured gold or silver [54]. We further note: (1) First, the transition to the diamagnetic state occurs at lower $T$ for higher $B$ (Figure 3c), which is consistent with the $B$-dependence of the resistive transition (Figure 3a). We find that $T_c$ decreases by nearly 12 K to as low as 216.5 K at 5 Tesla (Figure 3d). (2) Second, we also observe a strong repeatable noise in susceptibility for $T < T_c$, irrespective of $B$. The origin of this unique 'noise' remains uncertain, and further details can be found in Section S21 (SI).

To explore if the resistive and magnetic transitions in the NS films are concurrent, we subsequently measured the inductive response simultaneously with $R$ in several films with two-coil magnetometry as a function of $T$ (inset of Figure 4a, and Section S13, SI). The films were loaded on a ceramic holder equipped with spring-loaded electrical contacts and two coaxial coils (drive and sense), which allowed measurement of the $T$-dependence of $R$ and inductive response simultaneously (see Section S6, SI for more detail). In a superconducting transition, the onset of diamagnetism is expected to cause screening current that reduces the mutual inductance threading the drive and sense coils below $T_c$ [55]. Figure 4a shows the $T$-dependence of $R$ in Device P20519FEE_21, and that of the mutual inductance between the drive and sense coils. The real (inductive) and the imaginary (dissipative) components of the latter are denoted by $M'$ and $M''$, respectively. We observe that $R$ drops to the low resistance state at $T \approx 172.5$ K which is closely preceded by a decrease in $M'$ at $T \approx 165$ K (Figure 4b). The two-step decrease in $R$ is likely due to spatial inhomogeneity in nanoparticle density or electrode proximity effects (Section S14 – S18, SI). This may also explain the small difference (up to ~ 6 – 7 K, depending of device) in the $T_c$'s for the resistive and inductive transitions because $T_c$ close to the center, *i.e.* close to the coil axis, may also differ slightly from that at the peripheral contact regions (also See Section S17 for device P20319FEE_05).

From the structural characterization (Figure S1 and S2 in SI) of the nanoparticles, one estimates the Thouless temperature $T_{Th} = \frac{\hbar \pi^2 D}{k_B (2\pi r)^2} \sim 10^4$ K, where $D$ and $2r$ are the electronic diffusivity (~ 0.03 m²/s) and nanoparticle diameter ($\approx 1$ nm), respectively [47,49]. Although the nanoparticles may naturally host circulating persistent current because experimental $T \ll T_{Th}$, its maximum magnitude (~ $\frac{ev_F}{2\pi r} \approx 0.1$ mA) even in the ballistic limit is at least 100 times smaller than that observed in our experiment (Figure 2d and e). Emergence of a many-body coherence that couples the single nanoparticle persistent states, however, cannot be ruled out, which will presumably accompany a macroscopic magnetic transition as well.

Nonetheless, the close proximity of the resistive transition and that in inductive response may also suggest a superconducting transition, which is further supported by the change in $M''$ at the transition signifying dissipative coupling (Figure 4b). Assuming such a case, we estimated the in-plane London penetration depth ($\lambda$) [56,57] to be $\approx 800$ nm in Device P20319FEE_21 from the mutual inductance below the transition (see Section 7, SI). Notably, a reliable estimate of $\lambda$ is difficult here due to finite size effects arising from the inhomogeneous nature of the superconducting phase, as well as strong thickness variation across the film. Additionally, observation of resistive transition up to $B$ as large as 3 Tesla (or more) in Figure 3a suggests coherence length $\xi < 10 - 20$ nm, and thereby a type-II superconducting phase with Ginzburg-Landau parameter $\kappa > 50$.

We further studied the variation of $T_c$ as a function of the nanoparticle composition. In particular, Section S22 (SI) shows the transition in three samples containing different Au and Ag mole fractions. In each case the stoichiometric ratio has been altered by growing different amounts of Au over the same Au/Ag core NS. Figure 5a shows the variation of $T_c$ (corresponding to 95% reduction from the normal state resistance) with the Au mole fraction in these NS. It is evident that increased Au mole fraction on the NS significantly lowers $T_c$, although subsequent measurements showed that $T_c$ can also be influenced by aging, and repeated thermal cycles (Section S5, SI). Nonetheless, in view of this observation, we focused our attention towards NS with a low Au mole fraction (Section S1 (Materials and Methods), SI). Figure 5b shows the $T$-dependence of $R$ in film P21018FE0_02 (Table S1, SI) with low Au mole fraction ($x_{Au} = 0.63$). While a measurable transition could not be observed in the $T$ window ($\leq 350$ K) accessible to us, the $I - V$ characteristics, show no evidence of a voltage drop above the noise level of $\sim 100$ nV (inset of Figure 5b), even at the highest $T$. The corresponding upper limit of resistivity ($\sim 4 \times 10^{-11}$ $\Omega$.m) is again significantly lower than the bulk resistivity of highly conductive metals, and exhibits no dependence on $T$. Further, we observed that pellets of such samples are significantly diamagnetic ($\chi_v = -0.037$) under ambient conditions (Figure 5c), strongly suggesting the existence of a superconducting state at room temperature. The data shown in Figure 5c correspond to $x_{Au} = 0.73$. The diamagnetic character in such samples is sufficient to cause these to be visibly repelled by hand-held permanent magnets (also see Movie S1, SI).

In conclusion, we have carried out detailed electrical and magnetic characterization of nanocomposite films and pellets based on Ag nanoparticles embedded in an Au matrix. At the ambient pressure, films with Ag mole fraction $> 0.1$, and minimal environmental exposure, were found to undergo superconductor-like transition to vanishingly small electrical resistance ($R < 2$ $\mu\Omega$), corresponding to electrical resistivity $< 10^{-12}$ $\Omega$.m, below a critical temperature. The transition was observed at $T$ as high as 286 K, which could be increased beyond the experimentally accessible window ($\approx 350$ K) by tuning the nanoparticle density. SQUID magnetometry with the pellets of the same nanostructures indicates a strongly diamagnetic phase below the critical temperature. The concurrence of the resistive and magnetic transitions was further verified using two-coil magnetometry, which provides compelling evidence for a superconducting phase in our nanostructures that can be stabilized at the ambient conditions.

**Acknowledgements:**

AP and AG acknowledge financial support from the Indian Institute of Science. AP further acknowledges use of facilities created under the DST Nanomission grant (SR/NM/NS-1117/2012). SP acknowledges DST Swarnajayanti Fellowship. We further thank CENSE for access to their facilities. Additionally, we thank Prof. Naga Phani Aetukuri, Prof. T. V. Ramakrishnan and Prof. Vijay Shenoy for helpful discussions. DKT thanks Dr. Triloki Pandit for help with measurements.

**Author contribution:**

AP conceived the idea and designed the project. AP, DKT developed initial synthetic protocols, and carried out early measurements with help from SKS. SKS, BB, RM and DKT devised further protocols and prepared samples. SP contributed to device stabilization. AG, SI, PBS and TPS carried out electrical transport and inductive response measurements. AP and AG co-wrote the paper.

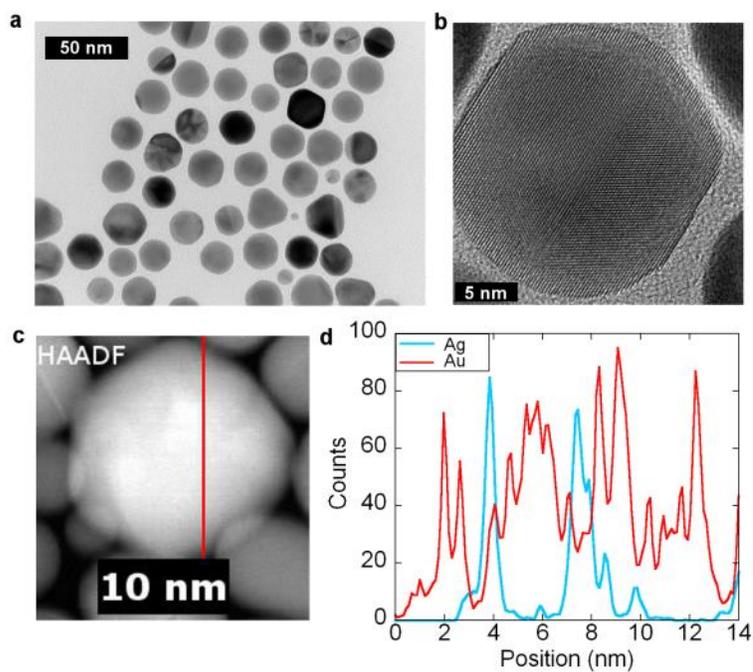

**Figure 1. Structural characterization of NS prepared in this work. (a)** TEM image of NS. **(b)** HRTEM image of a single NS. **(c)** HAADF-STEM map of a single NS. **(d)** The elemental distribution along the red line in panel (c).

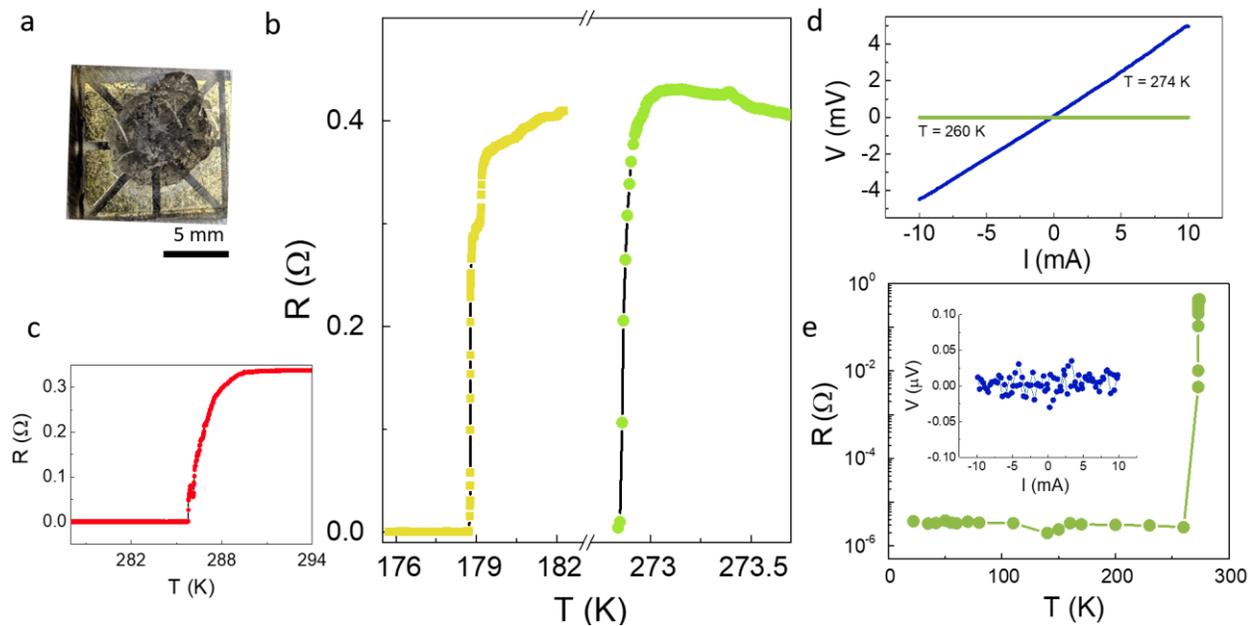

**Figure 2. Resistive transitions in nano-structured films.** (**a**) Micrograph of a typical nanostructures film on a patterned slide (**b**) Temperature-dependence of resistance in two representative samples, Device P20319FEE_06 (green circles), with a transition at $T_C = 272.85$ K, and Device P20519FEE_20 (yellow squares) with $T_C = 178.8$ K (Table S1 in SI). (**c**) Resistive transition in P20319FEE_06 following six days of ageing. The transition width has broadened while $T_c$ has increased to ≈ 286 K. (**d**) Current-voltage characteristics above and below the transition. (**e**) The resolution of resistance in the low resistance state (~2 µΩ), and the extent of resistance drop at the transition (nearly six decades), calculated from the slope of the $I - V$ curves as shown in the inset (See Section S7 of Supplementary Information).

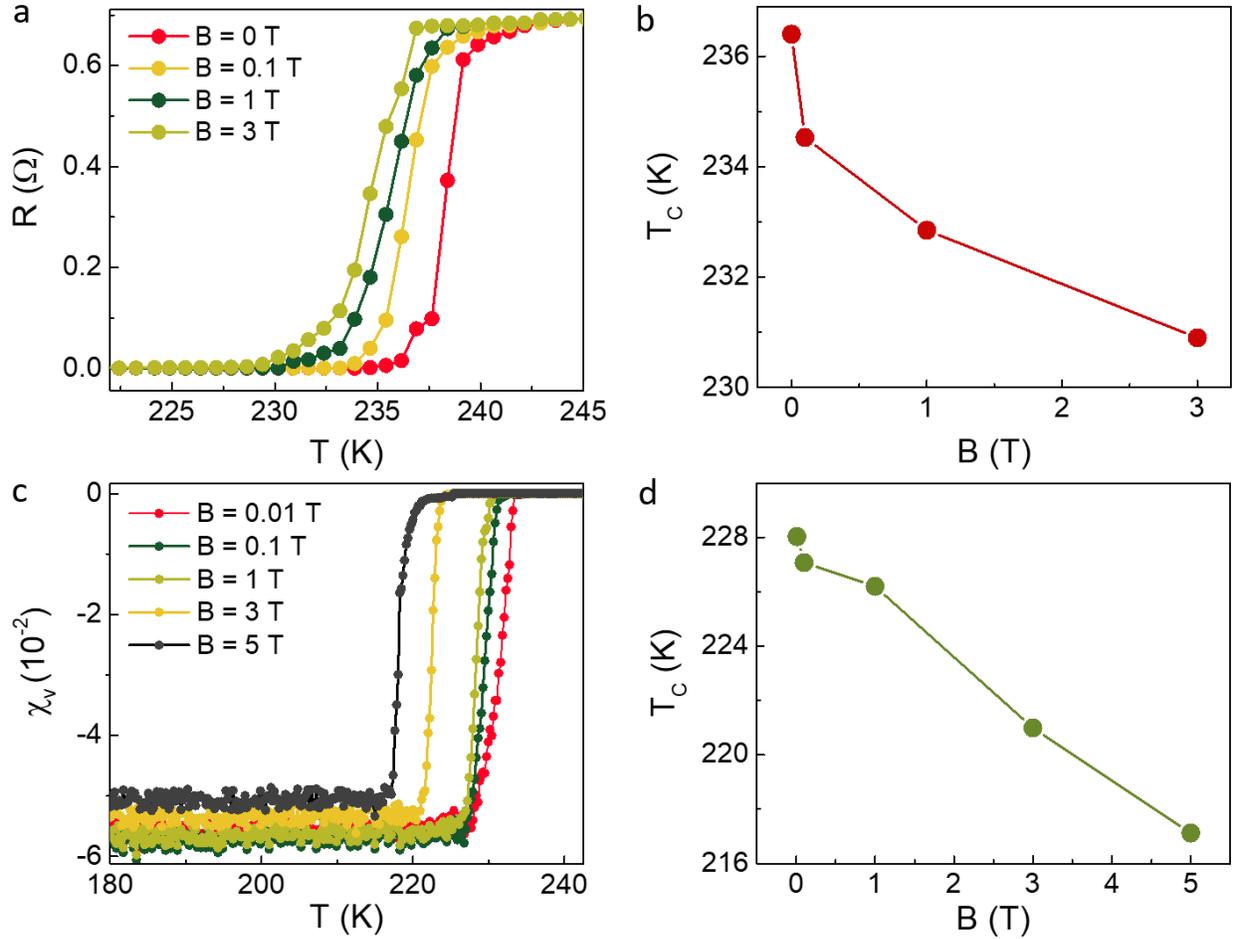

**Figure 3. Magnetic-field dependence of resistive and susceptibility transitions.** (a) Temperature-dependence of resistance of film P11017FE0_02 (Table 1, SI) at different external magnetic field. (b) Magnetic field-dependence of the characteristic temperature scale $T_c$ of the resistive transition. Here $T_c$ is evaluated at the 95% decrease from the normal-state resistance. (c) Variation of volume magnetic susceptibility of pellet P11117PE0_01 (Table 1, SI) with temperature at different external magnetic field. The diamagnetic state also exhibits larger fluctuations in susceptibility than that in the paramagnetic state (see Section S21, SI, for more detail). (d) Magnetic field-dependence of $T_c$, where $T_c$ is the temperature corresponding to 95 % of the final diamagnetic susceptibility.

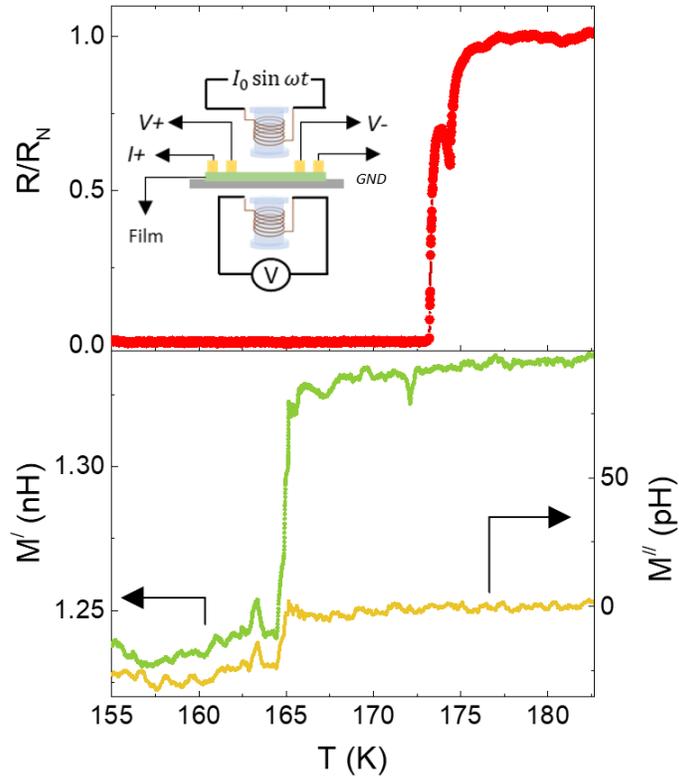

**Figure 4. Simultaneous transition in resistance and inductive response:** Simultaneous measurement of resistance (top) and inductive response (bottom) in Device P20519FEE_21 (Table S1, SI) as a function of temperature. The inset in the upper panel schematically shows the experimental setup used for electrical measurements. $M'$ and $M''$ are real and imaginary components of the mutual inductance between the drive and sense coils shown in the bottom panel.

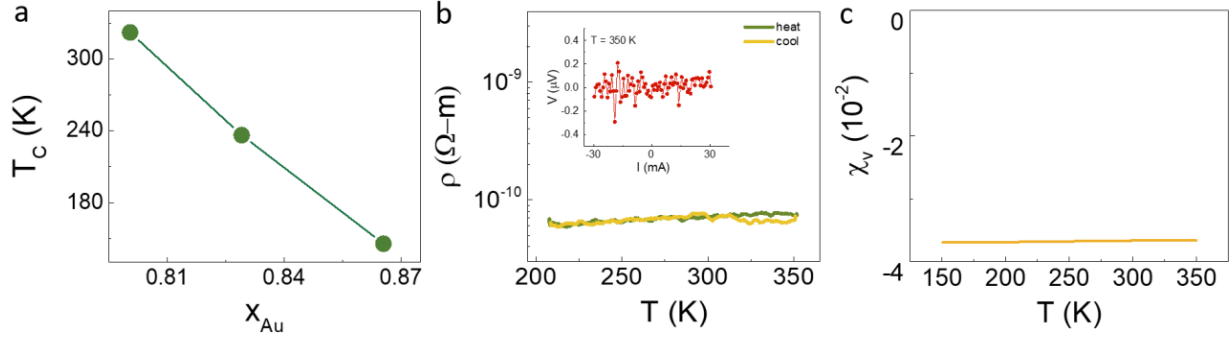

**Figure 5. Compositional tuning of superconducting transitions.** (**a**) Dependence of the resistive transition temperature $T_c$ on composition for a single starting nanostructure core. (**b**) Resistivity and (**c**) Susceptibility for NS with low Au mole fraction. Inset of (b) shows absence of voltage drop (within experimental error) up to biasing current of $\pm 30$ mA.

# Supplementary Information

# Coexistence of Diamagnetism and Vanishingly Small Electrical Resistance at Ambient Temperature and Pressure in Nanostructures

Dev Kumar Thapa[1], Saurav Islam[2], Subham Kumar Saha[1], Phanibhusan Singha Mahapatra[2], Biswajit Bhattacharyya[1], T. Phanindra Sai[2], Rekha Mahadevu[1], Satish Patil[1], Arindam Ghosh[2,3] and Anshu Pandey[1]*

**Affiliations:**

[1]Solid State and Structural Chemistry Unit, Indian Institute of Science, Bangalore 560012
[2]Department of Physics, Indian Institute of Science, Bangalore 560012
[3]Center for Nanoscience and Engineering, Indian Institute of Science, Bangalore 560012

*Correspondence to: anshup@iisc.ac.in

## Section S1. Materials and methods

### 1.1 Protocol 1:

**Materials:**

Hexadecyltrimethylammoniumbromide (CTAB) (≥98%) (Lot Nos. SLBL8064V, SLBN6918V, SLBT7250, SLBC8213V, SLBS1066V from Sigma Aldrich, Silver nitrate (ACS reagent, ≥ 99.0%) from Sigma-Aldrich, sodium borohydride powder from Sigma-Aldrich (≥98.0%), hydrogentetrachloroaurate(III)trihydrate (Lot. No. MKBT6878V and Batch No. 0000029562) from Sigma-Aldrich (ACS, 99.99% pure, metal basis) and Lot. No. AX0014 from Alfa-Aesar (ACS, 99.99% metal basis), L-Ascorbic acid (99%, Sigma Aldrich) were used as received without any further purification. All the aqueous solutions have been prepared in 18.2 M-Ohm milli-Q water to avoid contamination.

**Synthesis of Gold (Au) Nanosphere:**

Gold nanospheres of 10-12 nm size were synthesized by a seed mediated process. 5 mL of 0.5 mM $HAuCl_4$ was added to 5 mL of 0.1 M CTAB solution. The solution was stirred vigorously. To it was added 0.6 mL of 0.01 M $NaBH_4$. The final colour of the solution obtained was brownish indicating the formation of gold nanocrystals. These nanocrystals were used as seeds for the synthesis of 8-10 nm gold nanospheres.

A 500 mL growth solution comprising of 5 mM $HAuCl_4$, 0.1 M CTAB in 500 mL water was prepared. 3mL of 0.079 M Ascorbic acid. To the growth solution, 0.8 mL of Au seed was added. The solution was shaken well and kept for 1 hour.

**Synthesis of Superconducting Nanostructures (NS):**

Freshly prepared gold nanosphere solution was cleaned through centrifugation with water. The precipitate was re-dispersed in 10 mL of water. Then it was taken in a conical flask and 10 mL of 0.1 M CTAB solution was added to it. The solution was stirred. Next 5 mL of 1 mM silver nitrate solution was added rapidly into the solution and the timer for the reaction was switched on. When time reached 1 min 22 sec, 2 mL of 0.1 M $NaBH_4$ was added rapidly followed by drop wise addition of 100 microliters of 1 mM $HAuCl_4$ solution over 2 minutes. The solution was immediately cleaned and re-dispersed in water.

For experiments where a higher Au:Ag stoichiometry was desired, the ungrown sample was first cleaned twice with water through centrifugation (7392 RCF, 10 minutes). The obtained precipitate was then re-dispersed in 10 mL of 0.1 M CTAB solution. The solution was taken in 25 mL conical flask and the desired amount (typically 250 μL) of 1mM HAuCL$_4$ was added at 10 μL/3min. Before the addition of HAuCl$_4$ the solution was made reducing by adding NaBH$_4$ solution (2 mL 0.1 M). The reaction was stopped immediately after the addition of HAuCl$_4$ by centrifugation. In order to obtain the transition temperature at 148 K and 323 K, the growth was done by adding 425 $\mu L$ and 125 $\mu L$ of 1 mM HAuCl$_4$ respectively.

This procedure was found to work for lot nos. SLBL8064V, SLBN6918V, SLBT7250, SLBC8213V, SLBS1066V of CTAB, however failed with other grades and lots of CTAB. We could not identify the impurity/purity criteria that influence the efficacy of CTAB towards these reactions. In particular, it is evident from the $^1$H nuclear magnetic resonance spectra of that no NMR active impurities are present/absent in various successful and unsuccessful lots of CTAB. With a view to making the synthesis more stable and repeatable, we developed an alternate synthetic protocol that is presented below.

The below protocol was verified to work with the following sources of CTAB:

Batch No. 0000018772 from Sigma Aldrich (≥99% BioXtra)

Lot. No. SLBT7256 from Sigma Aldrich (≥98%)

Batch. No. E15A/1715/0804/13 (98%) S D Fine-Chem Limited

Batch. No. I16A/1816/1301/31 (99%) S D Fine-Chem Limited (SDFCL)

**1.2 Protocol 2:**

**Materials:**

Hexadecyltrimethylammonium bromide Batch No. 0000018772 from Sigma (≥99% BioXtra), Lot. No. SLBT7256 from Sigma Aldrich (≥98% pure), Batch. No. E15A/1715/0804/13 and I16A/1816/1301/31 from SDFCL. potassium iodide (221945-500G) Lot. No. STBH6776 from Sigma-Aldrich (ACS reagent, ≥99.0% pure), silver nitrate Lot.No. MKBT4516V from Sigma-Aldrich (ACS reagent, ≥ 99.0% pure) and Lot. No. MKBH9825V from Sigma-Aldrich (ACS reagent, ≥ 99.9999% trace metal basis), sodium borohydride powder Lot. No. STBH5482 from Sigma-Aldrich (≥98.0% pure), hydrogentetrachloroaurate(III)trihydrate Batch. No. 0000031810 and 0000040064 from Sigma-Aldrich (ACS, 99.99% pure, metal basis), L-Ascorbic acid (99%, Lot No. BCBX5254,

Sigma Aldrich) were used as received without any further purification. All the aqueous solutions have been prepared in milli-Q water to avoid any trace of metal contaminations.

**Synthesis of Gold (Au) Nanosphere:**

Gold nanospheres of 10-12 nm size were synthesized by a seed mediated process. 5 mL of 0.5 mM $HAuCl_4$ was added to 5 mL of 0.1 M CTAB solution. The solution was stirred vigorously. To it was added 0.6 mL of 0.01 M $NaBH_4$. The final color of the solution obtained was brownish indicating the formation of gold nanocrystals. These nanocrystals were used as seeds for the synthesis of 8-10 nm gold nanospheres.

A 500 mL growth solution comprising of 5 mM $HAuCl_4$, 0.1 M CTAB in 500 mL water was prepared. 3mL of 0.079 M Ascorbic acid. To the growth solution, 0.8 mL of Au seed was added. The solution was shaken well and kept for 1 hour.

**Synthesis of Superconducting Nanostructures (NS):**

In the first step, aqueous solutions of 0.1 M CTAB (5 mL), 0.1 M of KI and 1 mM silver nitrate (5 mL) were mixed. The mixture was continuously stirred for 4 min 30 seconds (solution 1).

In the second step, a separate vial with 5 mL aqueous solution of cleaned gold nanospheres and 2 mL aqueous solution of 0.1 M sodium borohydride were taken. The resulting solution was stirred continuously. 10 mL of solution 1 from above and 2 mL aqueous solution of 0.05 mM $HAuCl_4$ were added to the nano-sphere solutions over 8 minutes.

**Sample cleaning for pellets:**

Several batches of such samples were synthesized and mixed together. The sample was cleaned through centrifugation with water. The centrifugation was done five times. The obtained precipitate was collected in 30 ml vial and was kept for drying. To the dried sample, 5 mL of $CHCl_3$ (Chloroform, LR grade) was added and the solution was sonicated for 10 min. The solution was kept in $CHCl_3$ solution for 4 hrs. After 4 hours the solid sample was separated from $CHCl_3$ through centrifugation and a fresh $CHCl_3$ was added. The above process of sonication and the addition of $CHCl_3$ after every four hours were carried out for 2 days. On the next step, the sample in $CHCl_3$ is centrifuged and precipitated. The precipitated was kept for drying. Once dried the solid sample was washed with acetone several times and left for drying. To the dried sample 5 mL of > 1 M KOH (Potassium hydroxide) solution in water was added. The sample in KOH solution was sonicated for 5 minutes and was kept for half an hour. KOH solution was changed every 30 minutes and the sonication step was repeated. The process was continued over 24 hours. The final obtained mass of fine grains was pressed to form a pellet.

**TEM characterization:** Transmission Electron Microscope (TEM) grid was prepared by depositing samples from aqueous solution. HR-TEM and STEM images were obtained on a Themis TITAN transmission electron microscope. STEM-EDX elemental mapping was also performed using the same instrument.

**Magnetic susceptibility measurement:** Pellets were prepared by pressing grains isolated above in a titanium die. The typical pellet weight for magnetometry measurements was about 65 mg. The magnetometry measurement was done in a SQUID (MPMS®3, Quantum Design). The sample was filled into a tube that was then attached to the sample holder.

**Films casting procedure**

**Method 1:** As synthesized samples were cleaned twice by centrifugation. The obtained precipitate was re-dissolved in water and the solution was then drop cast on a glass substrate which had six gold metallic pads (100 nm height with and an equidistant separation of 100 µm) deposited on it. The film making was carried out inside a glove box. The dried film was washed with propanol once. To the dried film, $CHCl_3$ (A.R grade, SDFCL) was added and left for drying. Following this, the film was dipped into 2 M aqueous KOH (A.R grade, SDFCL) and left for drying. The film was then gently washed with methanol and propanol. Care was taken so as not to wash away the nanoparticles. The process of washing was repeated twice. The process of addition of $CHCl_3$ followed by the addition of KOH was repeated. The next round of sample was drop-cast on top of this film and crosslinking was again carried out. This process was repeated multiple times (typically 15) until the desired density of film is obtained. Films prepared prior to August 2018 employed this method.

**Method 2:** The pristine sample was cleaned twice by centrifugation. Films were prepared inside a glove box. The obtained precipitate was re-dissolved in water and the solution was then drop-cast on a glass substrate with eight gold pads in Van der Pauw configuration (Details describes in Section 6). The film was treated with >5 M KI (ACS reagent ≥ 99.0%) aqueous solution. The film was then washed with water several times and subsequently treated with $CHCl_3$ (A.R. grade, SDFCL) followed by dipping into 3 M aqueous KOH (A.R.

grade, SDFCL) for 10 minutes. The process of addition of CHCl$_3$ followed by the dipping was repeated twice. After each addition the film was washed with water several times. The next round of sample was drop-casted on top of this film and crosslinking was again carried out. This process was repeated several times until the desired quality of film is obtained.

**Resistance and inductance measurements:**

For samples where resistance measurements were performed in AC, an SRS 830 lock-in amplifier was used, with an AC excitation current of 1 mA. For DC-resistance measurements, a Keithley 2400 source-meter was used to pass a current of 1 mA, and the voltage was measured with a Keithley 2002 multimeter.

For inductive response measurements in the two-coil geometry, an AC-current of 1 mA was passed through the drive coil at a fixed frequency $f = 43.33$ KHz, using an SRS 830 lock-in amplifier. The phase of the lock-in amplifier was kept fixed at 90 degrees during the phase-locked detection of the voltage developed across the sense coil. The winding on both coils were carried out in dipolar manner.

**Section S2. AFM image of Ag particles**

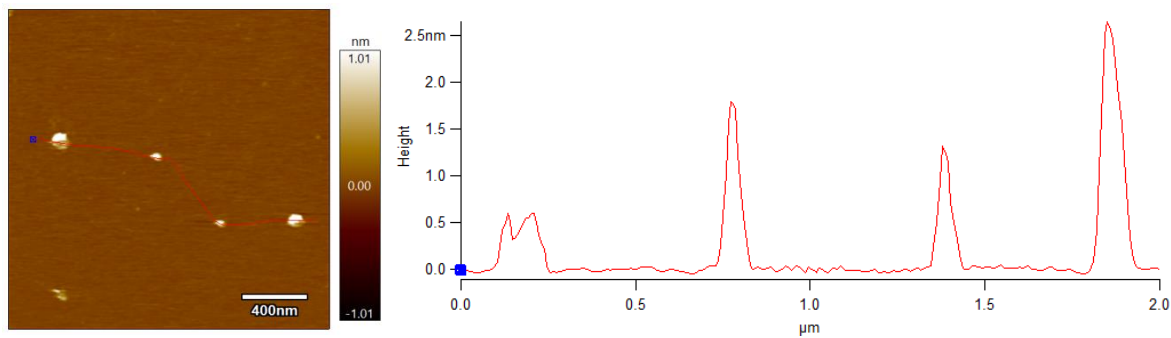

**Figure S1. AFM image of Ag nanoparticles.** AFM images of Ag nanoparticles deposited onto a mica substrate. The clusters were identified by verifying the topography of blank mica. An Asylum Cypher ES AFM was used to acquire the image in tapping mode (AC Air topography). The lateral dimensions of particles are convoluted with the tip dimensions as well as measurement artefacts arising from scan rates. The height profile is consistent with ~1 nm silver clusters.

Ag nanoparticles were prepared by following Protocol 2 as described in the supporting Section S1 without the addition of Au nanospheres and HAuCl$_4$ solution. The synthesized clusters were deposited onto a substrate by drop casting. The drop cast solution was allowed to stand for 10 minutes. It was subsequently washed off with water. The rinsed substrate was dried.

**Section S3. Structural characteristics of the NSs.**

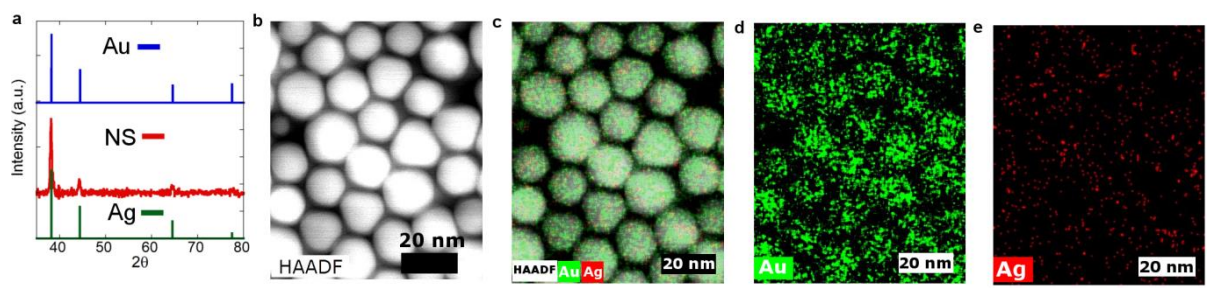

**Figure S2. Structural characteristics of the NS. (a)** Powder XRD pattern of NS (red) matches with the standard pattern of Au (blue). and Ag (green). **(b) - (e)** HAADF-STEM of the NS and the corresponding elemental mapping.

Powder XRD patterns were collected using a 0.154 nm source. Au and Ag have similar lattice constants and as a result, their independent contributions cannot be resolved within the NS. We therefore turned to microscopy to characterize the distribution of Au and Ag in the NSs. This is exemplified in the structural and elemental mapping shown in panels b-e. These data are consistent with a structure comprising of Ag nanoparticles distributed within an Au matrix.

**Section S4. Energy Dispersive X-Ray spectrum and elemental composition of the NS.**

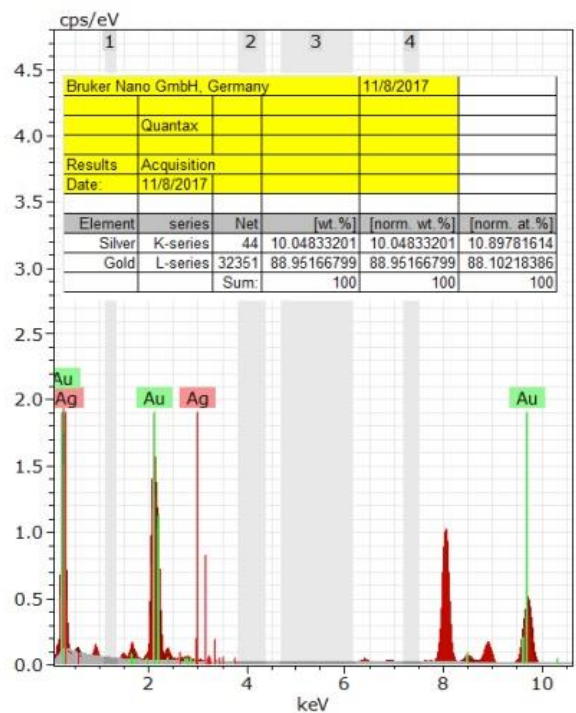

**Figure S3.** Energy Dispersive X-Ray spectrum and elemental composition of the NS.

The composition of the NS was analysed through EDAX measurements. The results on a typical structure are exemplified in Figure S3 above that shows a NS with $x_{Au}$ of 0.88. This paper reports results on samples with $x_{Au}$ greater than 0.5.

**Section S5.1. TEM and HRTEM images of NS**

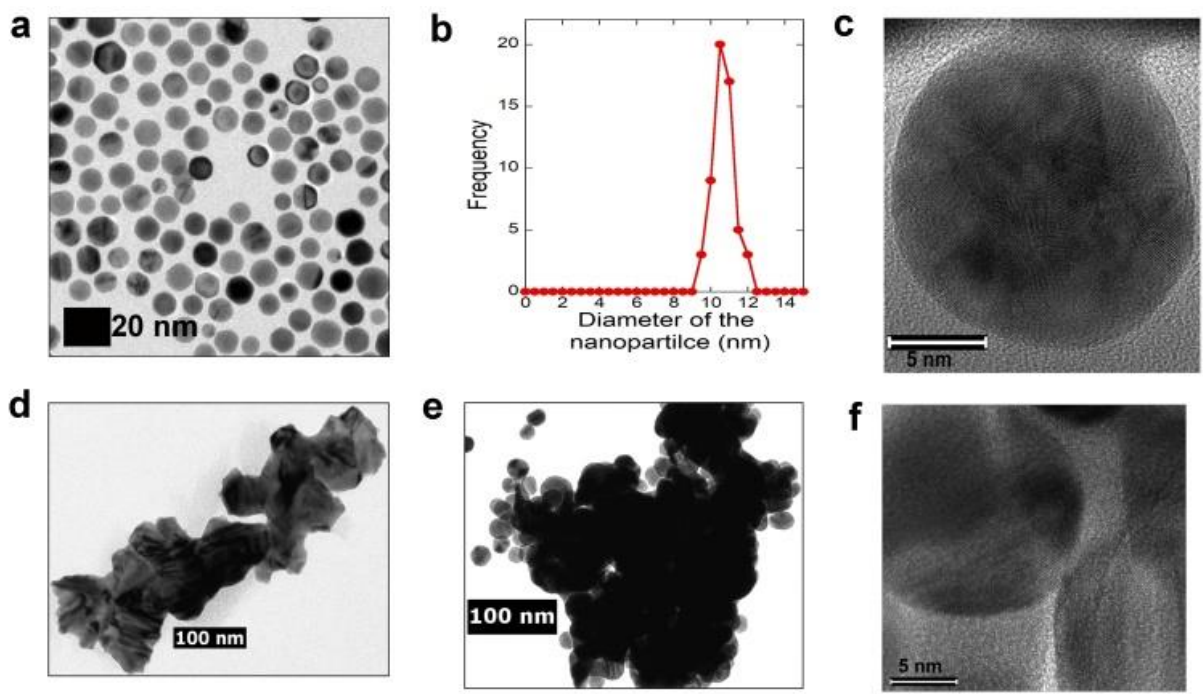

**Figure S4. TEM and HRTEM images of NS (a)** TEM images of NS. **(b)** Size distribution corresponding to panel (a). **(c)** HRTEM image of the NS. **(d)-(e)** TEM images of agglomerated NS and **(f)** HRTEM image of agglomerated NS. To acquire **(d)**-**(e)** the sintered NS were sonicated in water for 2 hours, the dispersed particles were then drop cast onto a carbon coated copper grid for microscopic imaging.

The morphological characteristics of NS and NS aggregates are shown in Figure S4. Figure S4a shows a TEM image of an ensemble of NS. The corresponding size distribution is shown in Figure S4b, while Figure S4c shows a corresponding HRTEM image. Chemical sintering protocols used in pellet making as outlined in Section S1 lead to agglomerates shown in Figure S4d-f.

## Section S5.2. AFM image a deposited NS film

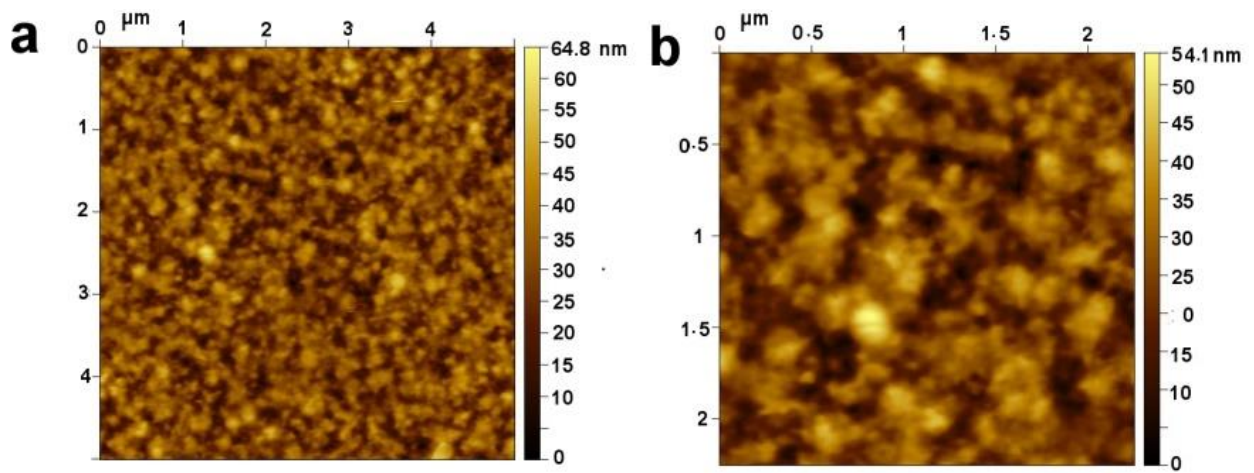

**Figure S5.** AFM image a deposited NS film used to study electrical characteristics.

The profile of a film of NS used for measurements is shown above. The above images correspond to a sample prepared using protocol 1 described in Section S1.

# Section S6. Details of the fixture for simultaneous measurement of electrical resistance and inductive response

To reduce the time of exposure to ambient conditions, and simultaneous investigation of both resistivity and any associated magnetic transitions, specially designed sample holders, compatible with the Van der Pauw geometry of the contacts pads were constructed. The holder was fabricated out of macor, which has a thermal conductivity of 1.46 W/m.K. Eight holes were drilled and pogo pins with a base diameter of 1.07 mm, were attached through the holes, which contacted the Cr/Au pads to ensure electrical continuity. The macor sample holder is composed of two parts: the top part, contained both the pogo pins and the drive coil, and was firmly anchored to a copper block to ensure proper thermalization. While one end of the spring-loaded pogo pins touched the Cr/Au contacts, the other end was soldered to a printed circuit board attached to the cryostat using 100 micron copper wires. The bottom part contained a step of 0.6 mm approximately to press the glass slide against the pogo pins. This part also contained the sense coil for mutual inductance measurements (also see Section S13).

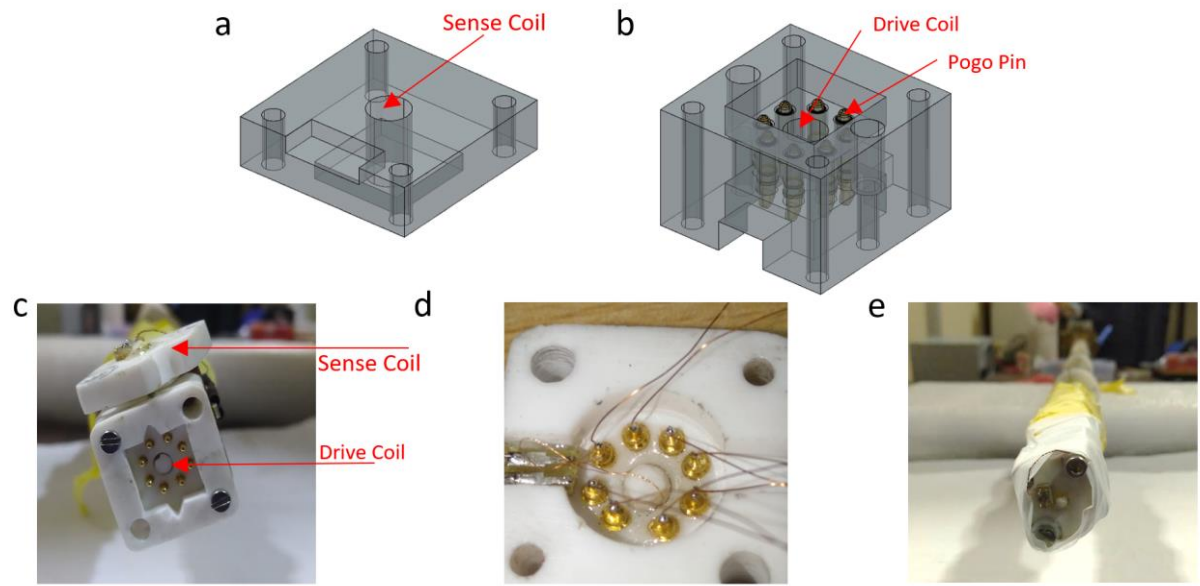

**Figure S6. Design of the sample holder.** (a) Design of the top part of the sample holder, containing the sense coil. (b) Design of the bottom part of the sample holder, containing the drive coil. This part contains 8 pogo pins, which electrically connect the sample to the instrument. (c) Optical micrograph of the actual sample holder containing the drive coil, and the pogo pins. (d) Optical micrograph of opposite side of the bottom part, where copper wires are soldered, and connected to a PCB mounted on the cryostat. (e) The sample holder assembly, after the sample has been loaded. The reasonably high thermal conductivity of macor, and strong thermal coupling of the pins to a copper base allowed us to reduce the temperature difference between the sample and the thermometer (Pt-100 resistor anchored on the copper base) to less than 0.1 K.

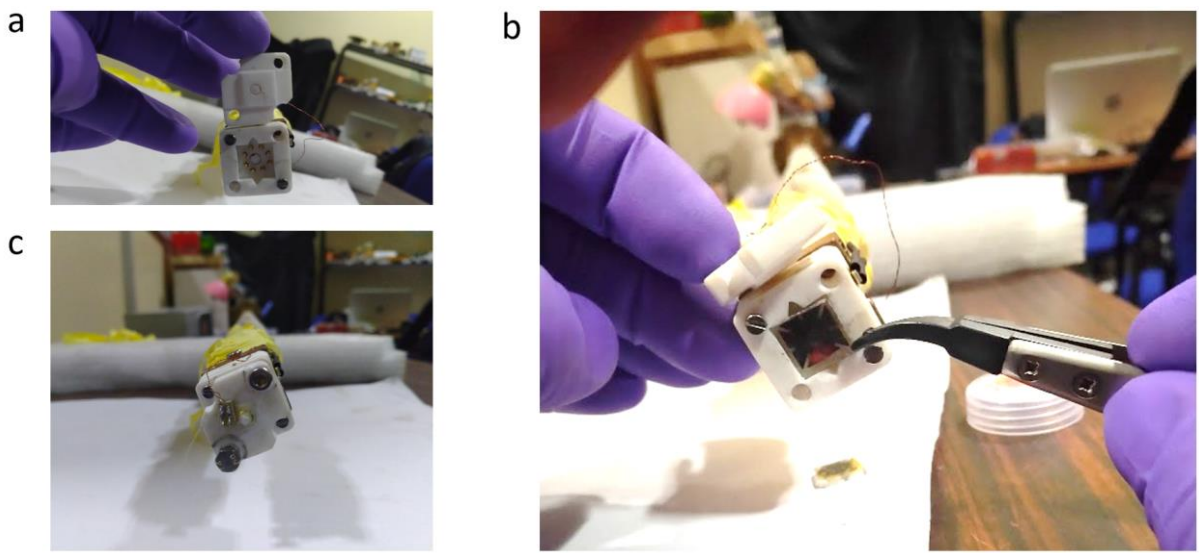

**Figure S7. Sample loading process.** (a)-(c) The film on the glass slide is inserted into the grove in the bottom part. The top part of the macor sample holder is then pressed against the sample, and the two parts are screwed together. No separate bonding or soldering of the contact pads is required in this process. The main advantage is that the loading of the sample onto to the measurement unit can be achieved within a few seconds.

**Van der Pauw configuration of contact pads**

For film-type samples, 5/50 nm Cr/Au was deposited on standard glass slides, which were covered with shadow masks with Van der Pauw configuration, fabricated from 50-micron stainless steel sheets. The glass slides were then cut into 10 mm by 10 mm pieces.

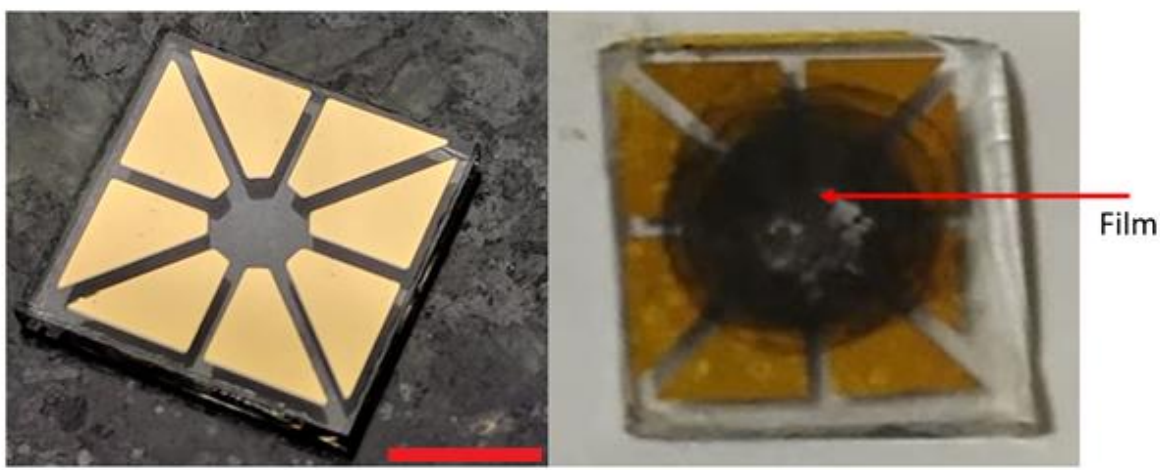

**Figure S8. Configuration of contact pads:** (Left panel) Typical optical micrograph of a glass slide (10 mm by 10 mm) after deposition of Cr/Au electrodes. The red scale bar is 5 mm in length. (Right panel) Optical image of the film after it was drop-cast on the glass slide.

**Section S7. Evaluation of uncertainty in resistance measurement in the zero-resistance state**

The DC measurements were performed with Keithley 2400 source meter which was used to source the current ($I$), while the voltage was measured using Keithley 2002 multimeter. To evaluate the resistance limit in the superconducting state, at first, $I$ was swept between $\pm 10$ mA in steps of 50 µA, and the voltage was recorded. Between 25 to 50 sweeps were recorded, following which the average voltage at each $I$ was calculated, finally followed by a sequential 10-point averaging. The averaged voltage values form a histogram with a Gaussian distribution, from which the full width at half maximum (FWHM) was extracted. The resistance limit is then obtained by dividing the FWHM by the maximum current (10 mA). The resistance limit calculated was $\sim 2$ µΩ, for most of our measurements.

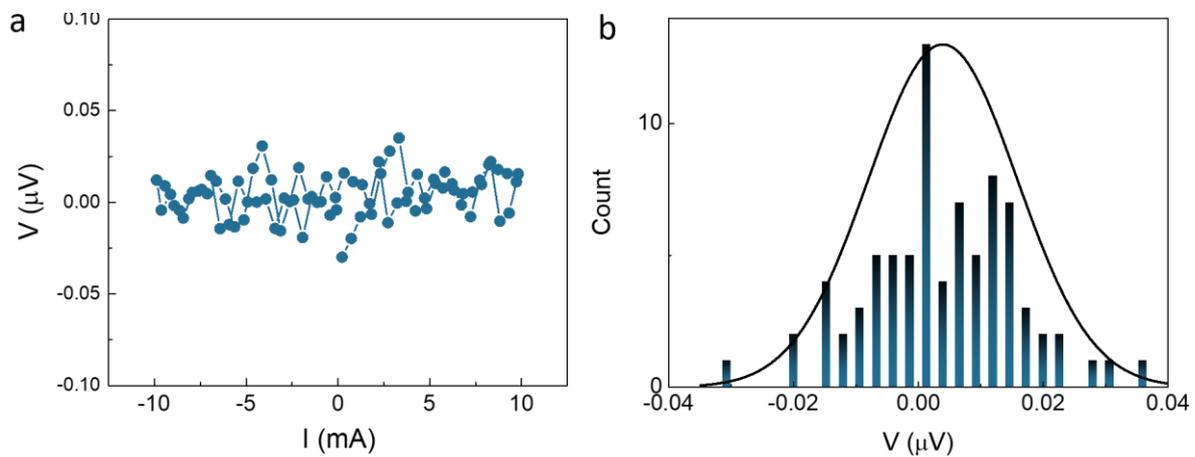

**Figure S9. Evaluation of resistance uncertainty.** (a) Average value of voltage vs. current in sample P20319FEE_06. (b) Histogram of the measured voltage value in Figure S9(a). The solid black line is the Gaussian fit, scaled to the maximum value.

**Section S8. Anisotropy in resistance**

The nature of resistance vs. temperature of the samples along different channels were often found to be different. This is possibly due to inhomogeneity in the cluster density or environmental contamination in the sample. As an example, in sample P20119FES_07 in a particular thermal cycle, only one of the channel's showed a drop in *R*, while the other channels were metallic. In sample, P20119FE0_08, two of the channels showed metallic behaviour, while another channel displayed insulating behaviour.

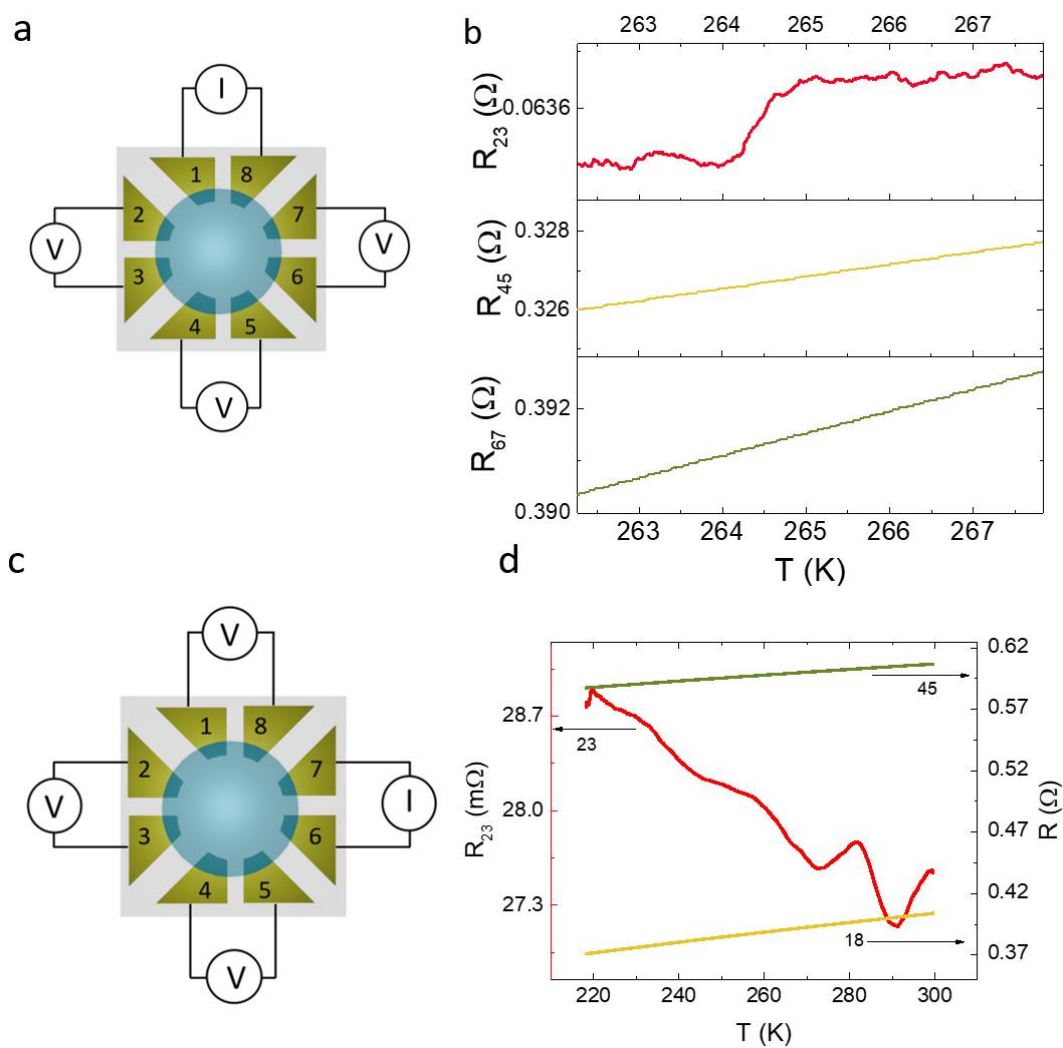

**Figure S10. Anisotropy in resistance.** (a) Schematic illustration of the contact combinations used for current and voltage measurements. (b) Resistance vs temperature of different channels of P20119FE0_07. While one of the channel's showed a drop, two different simultaneously measured channels show metallic behaviour. (c) Schematic illustration of the contact combinations used for current and voltage measurements in sample P20119FE0_08. (d) Resistance vs temperature of different channels. While two of the channels show metallic behaviour, one channel (2-3) shown insulating behaviour.

## Section S9. Effect of thermal cycle in a highly resistive film

The temperature-dependence and effect of thermal cycle in a highly resistive film is shown in Figure S11. The sample shows a metal to insulator transition in some thermal cycles, and a large hysteresis as well.

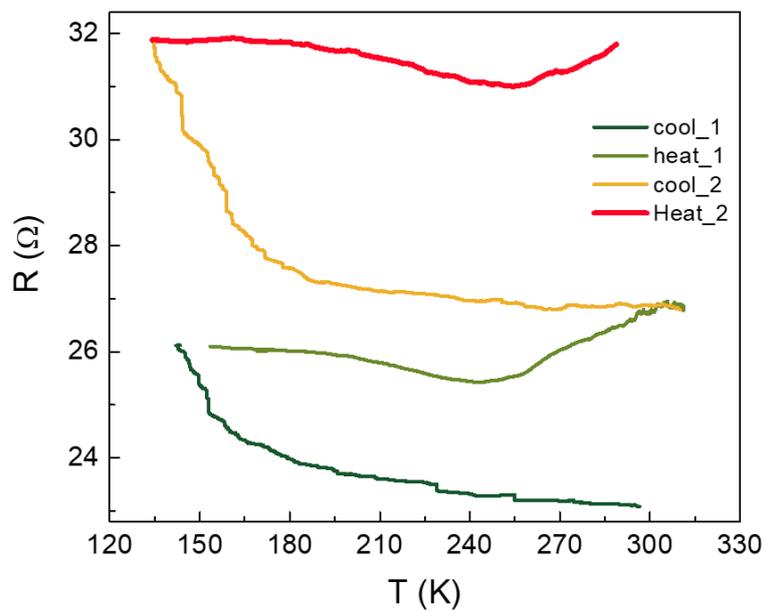

**Figure S11. Effect of thermal cycle on sample P20918FE0_01.** Resistance vs temperature for several thermal cycles, showing insulating or metal to insulator transitions in different cycles.

## Section S10. Electrical continuity between contacts in the superconducting and the normal states

In order to ensure the electrical contacts used in the measurement of the superconducting state maintained metallic electrical connectivity through the transition, we measured two-probe $I-V$ characteristics among the relevant contacts in a pair-wise manner. The superconducting transition in Figure 2b (main text) was obtained with contacts 2, 3, 6 and 7, with contacts 6 and 7 being used as voltage probes, while 2 and 3 employed as the current leads. The two-probe $I-V$ traces between different contacts at $T = 170$ K (below transition) and 322 K (above the transition) are shown in Figure S12. All $I-V$ characteristics show linear behaviour, with resistance ranging between $10-150$ Ω, indicating that different regions of the sample are not only electrically connected, but the resistance is that expected of a good metal.

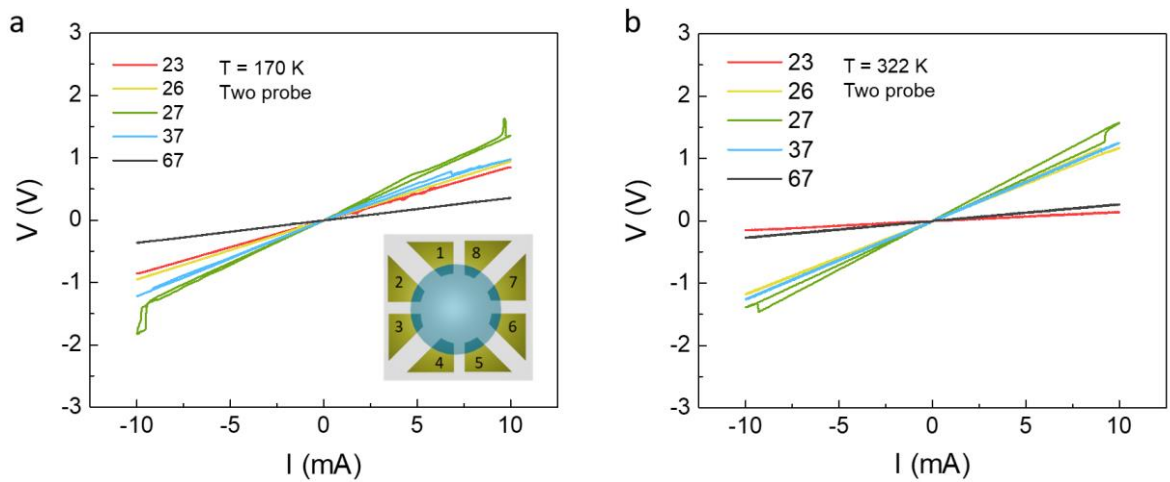

**Figure S12. Two probe I-V characteristics.** (a) Two-probe I-V characteristics between different contact combinations at $T = 170$ K, when the sample was in the superconducting state. (b) Two-probe IV between different contact combinations at $T = 322$ K, when the sample was in the normal state.

### S10.1. Reciprocity of the superconducting behaviour

In addition to metallic connectivity we have also confirmed the reciprocity of the superconducting behaviour by swapping the current and the voltage leads. Figure S13 below shows that the superconducting state is robust against the interchanging of the leads.

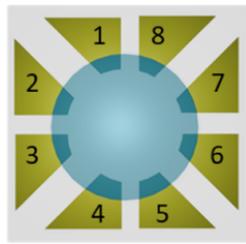 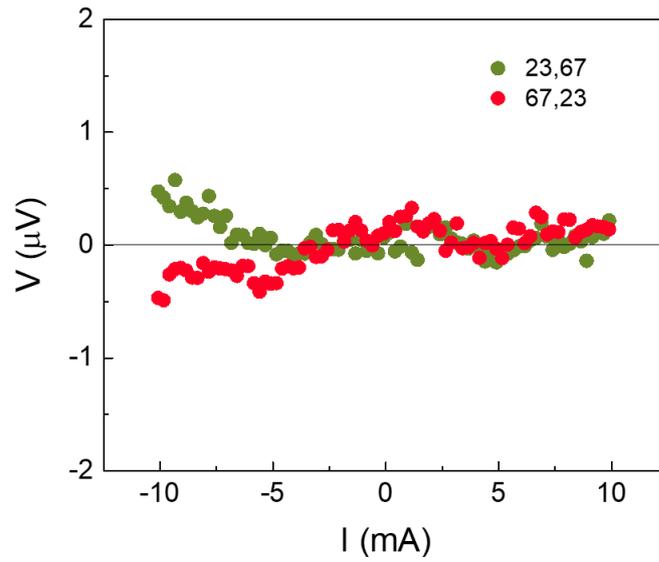

**Figure S13. Four probe $I - V$ characteristics.** Here first and second numbers indicate current probes, while the third and fourth numbers indicate voltage probes.

**Section S11.1. Ageing effect.**

The nature of superconducting transition was found to be influenced by ageing of the film. This is illustrated with device P20319FEE_06 in Figure S14, where the left and the right panels were recorded six days apart. It is to be noted that we carried out continuous electrical measurements (four-probe $I-V$ characteristics) and subjected the film to multiple thermal cycles during this period.

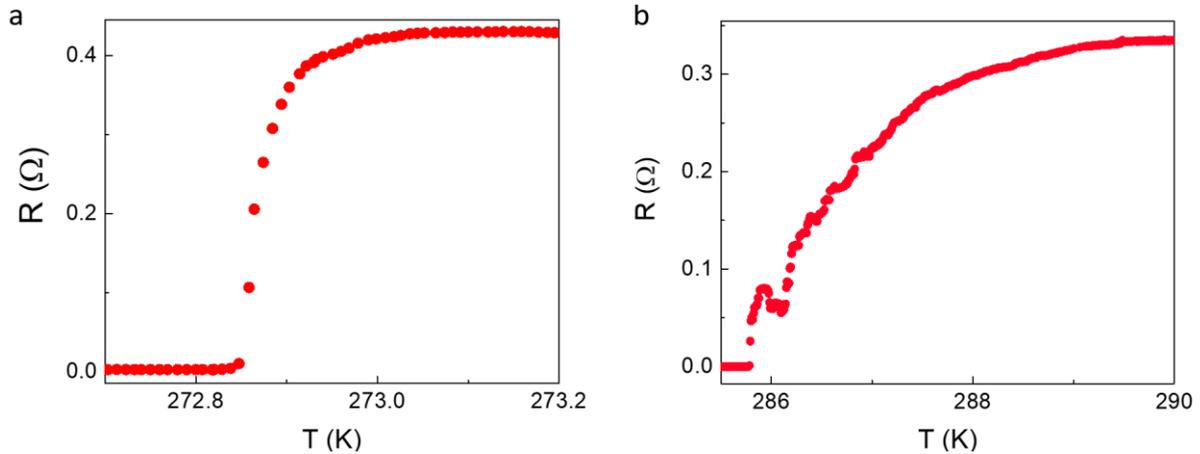

**Figure S14. Ageing effect in P20319FEE_06.** (a) The superconducting transition in the pristine sample. (b) The superconducting transition after six days. The film stopped working soon after due to contact failure.

Two effects of ageing are noteworthy:

1. The transition is considerably broadened, and

2. The transition temperature has increased by nearly 14 K.

**Section S11.2. Additional effects of ageing and thermal cycles**

The nature of transition observed in different thermal cycles in P20319FEE_06 is shown in Figure S15. While the general trend seems to be broadening of the transition with increasing number of thermal cycles, we also observe fluctuations between the normal and the superconducting states (or "re-entrant superconductivity") in one of the cycles. (See Section S15 for more discussion on this observation.)

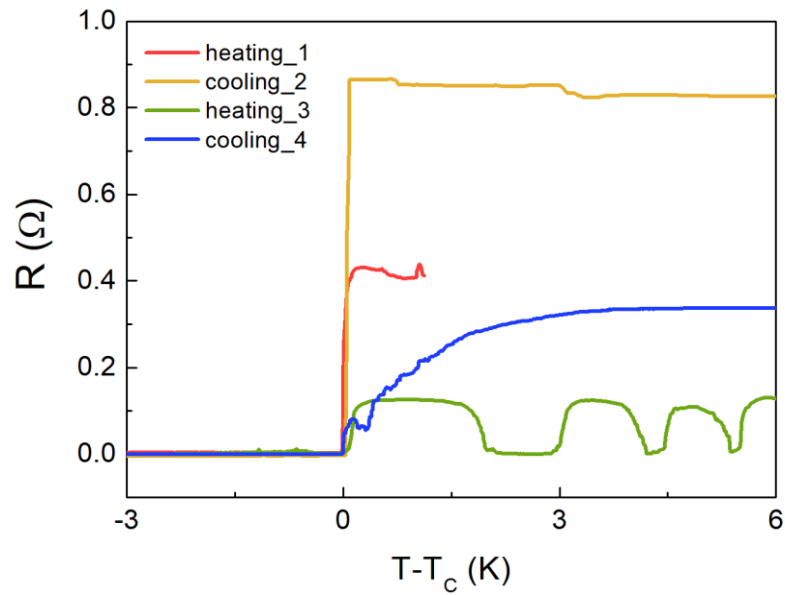

**Figure S15. Effect of thermal cycle on the transition in sample.** The superconducting transition in sample P20319FEE_06 in different thermal cycles. The temperature-axis is shown in terms of $T - T_C$ to highlight the change in the nature of the transition.

**Section S12. Stability of the zero-resistance state**

The zero-resistance resistance state in sample P20319FEE_06 was stable for over 6 days (Section S11.1). During this period, we measured the $I-V$ characteristics, which is presented in Figure S16 below. It should also be noted that the films were also subjected to continuous heating at 1.6 K/hour from $T = 160$ K, culminating in superconductor to normal metal transition at $T = 272.85$ K.

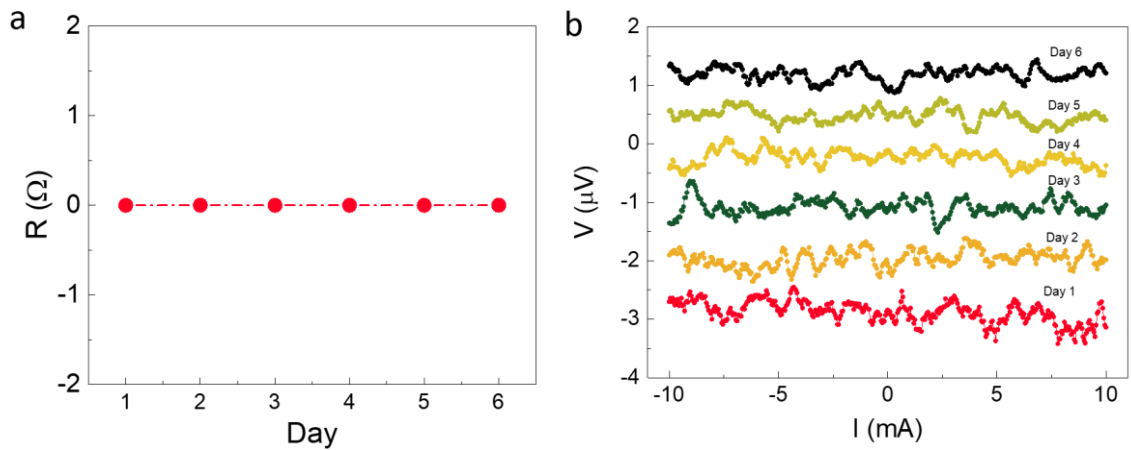

**Figure S16. Stability of the superconducting state.** (a) The superconducting state was stable for over 6 days, during which continuous electrical measurements were being performed. (b) The $I-V$ characteristics at the corresponding temperatures. The traces have been vertically offset for clarity.

## Section S13. Two-coil method for susceptibility measurements

To measure the inductive response of our films during the resistive transition, a two-coil method, which is based on Faraday's law of magnetic induction was used. According to Faraday's law, any change in magnetic flux will generate an electro motive force in a coil. The basic method to determine any magnetic transition is given below:

The primary coil with $N_P$ turns is excited with an AC-current source while the induced voltage is measured across the secondary coil, having $N_S$ turns. The coils are assumed to be cylindrical with a diameter of $R$, and the centre to centre separation between the coils is $d$. The magnetic field produced by the primary coil due to a fixed current ($I$) at the centre of the secondary coil can be calculated using Biot-Savart's law, and can be expressed as (for $d \gg R$)

$$B = \frac{\mu_0}{2\pi} \frac{N_S A}{d^3} I \qquad (S1)$$

where $A = \pi R^2$ is the average area of each turn of the coil. The induced voltage ($V$) in the secondary coil due to a time-varying current in the primary is given as

$$V = -N_p \frac{d\varphi}{dt} \qquad (S2)$$

$$V = -N_p A \frac{dB}{dt} \qquad (S3)$$

$$V = -\frac{\mu_0}{2\pi} \frac{N_S N_p A^2}{d^3} \frac{dI}{dt} \qquad (S4)$$

Here $\varphi = BA$ is the total flux through one turn of the sense coil. The pre-factor of $dI/dt$ on the right hand side of equation S4, is effectively the mutual inductance $M$ between the coils, and parameter of interest here.

Any change in the magnetic state of the sample, especially any emerging diamagnetic behaviour in the context of superconductivity, would screen the magnetic field generated by the primary coils due to screening current in the sample. The degree of screening is dependent on the magnetic penetration depth of the sample. This would lead to a reduction in $M$, which can be readily measured.

In our case, the primary (drive) coil with 36 turns (3 layers with 12 turns in each layer), which is on the upper macor piece, is connected to an AC-source (output of SRS 830 lock-in amplifier). The secondary (sense) coil with 48 turns (4 layers with 12 turns in each layer) is attached to the lower macor fixture. The diameter of the coils was 1.51 mm.

## Section S13.1. Control sample: Pb film

We have carried out control experiment of simultaneous measurement of the resistance transition and inductive response with a 120 nm thick Pb film, deposited thermally on a glass slide on 5/50 nm Cr/Au in Van der Pauw geometry (inset of Figure S17). The purpose of the control experiment is two-fold: (a) demonstrate simultaneous resistive and inductive transition in a conventional well-characterized superconductor, and (b) calibrate the two-coil set-up (including evaluation of the spurious pick-up voltage etc.).

As can be seen in Figure S17, the sample showed a transition at a temperature $T = 7.63$ K, consistent with values reported in literature. While the component of the sense coil voltage $V_x$, which is detected at 90° to the excitation current to determine the real component of $M$, showed a reduction in magnitude, the quadrature component $V_y$ showed a dip, representing the dissipative or resistive coupling during superconducting transitions. The pickup voltage $V_x$ below the transition temperature was identified as the background due to spurious inductive coupling, and subtracted from the experimental $V_x$ in subsequent measurements.

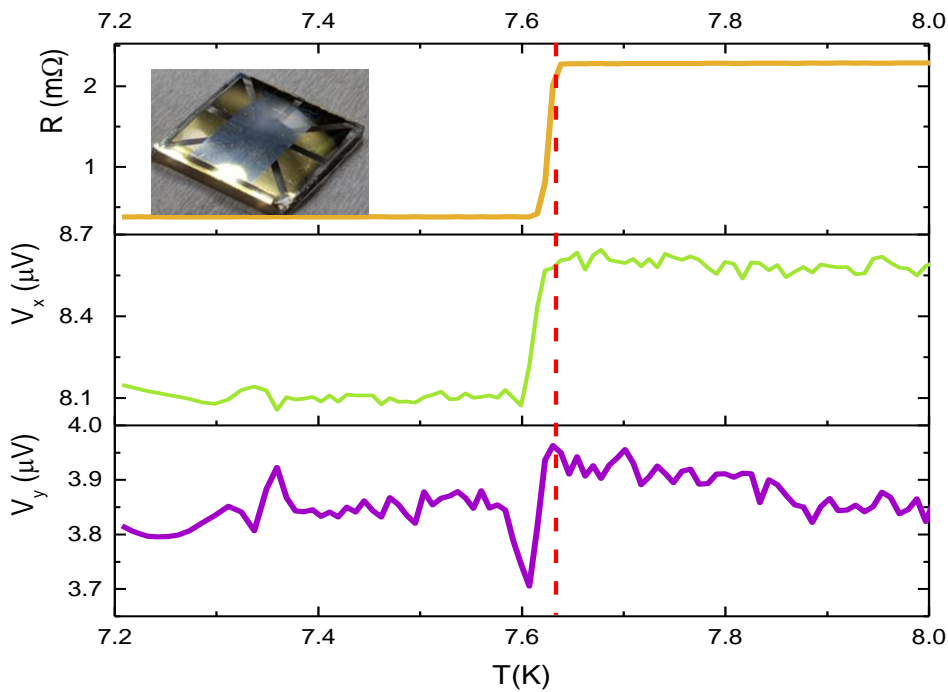

**Figure S17. Simultaneous measurement of resistance and inductive response in Pb film across superconducting transition.** (a) Resistance vs. temperature for the lead film of thickness 120 nm. (b) Temperature dependence the component $V_x$ of the pick-up voltage detected at 90° to the drive coil excitation current, that represents the real part ($M'$) of the mutual inductance between the drive and sense coils. $M'$ shows a clear drop at the same temperature, where the resistive transition occurs. (c) The quadrature component $V_y$ of the pick-up voltage, showing a narrow dip at the transition.

## Section S13.2 Calculation of penetration depth

The penetration depth has been estimated using expression valid for an infinite film in the thin limit, which is given as (1-3)

$$M = \frac{\lambda}{\sinh\left(\frac{d_f}{\lambda}\right)} \left[2\pi\mu_0 \frac{r_1 r_2}{h^2} \int_0^\infty dx\, x\, e^{-x} J_I\left(x\frac{r_1}{h}\right) J_I\left(x\frac{r_2}{h}\right)\right] \qquad (S5)$$

Here $M$ is the measured mutual inductance in presence of the film, $d_f$ is the film thickness, $\lambda$ is the magnetic penetration depth, $h$ is the centre to centre distance between the coils, $\mu_0$ is the magnetic permeability, and $J_1$ is the first order Bessel function of the first kind. The term in the square brackets in the right-hand side of the equation contains the information about the coil geometry and is calculated numerically. For multiple turns, the integral is summed over all possible coil configurations.

Figure S18 below shows the temperature-dependence of the (background-corrected) pick-up voltage $V_x$ (measured at 90° phase with respect to the AC drive current) and its quadrature component $V_y$ in device P20319FEE_05 (also see Figure S29 in SI ). The mutual inductance $M$ between the drive and sense coils is evaluated from $M = V_x/2\pi f I$, where $f = 43.3$ kHz and $I$ are the drive current frequency and rms magnitude of the drive current, respectively. Hence the drop in $V_x$ at ≈ 192 K, indicate a reduction in the $M$, that can be attributed to the diamagnetic screening in the superconducting state.

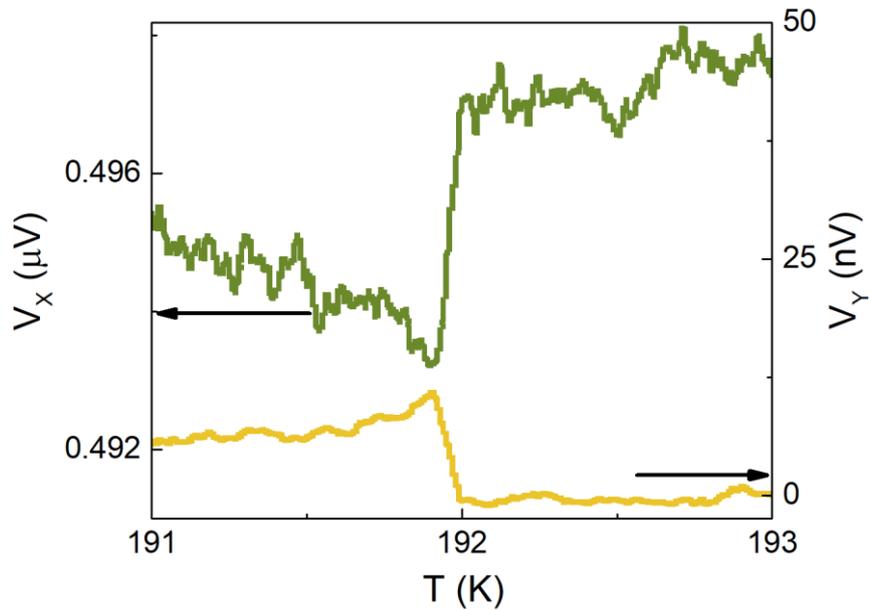

**Figure S18. Sense coil voltage across transition.** Temperature-dependence of the voltage (in-phase and quadrature) across the sense coil in device P20319FEE05.

**Section S14. Incomplete transitions in resistivity**

While most of the samples measured were metallic in nature (See Table SI), some of the samples showed incomplete transitions. Such a transition is illustrated in Figure S19 with P20918FE0_02 at $T = 176$ K. The drop-in resistance was $\Delta R = 68$ %. The sample recovered to its normal state resistance after the drop. This behaviour was repeated in the next heating cycle as well.

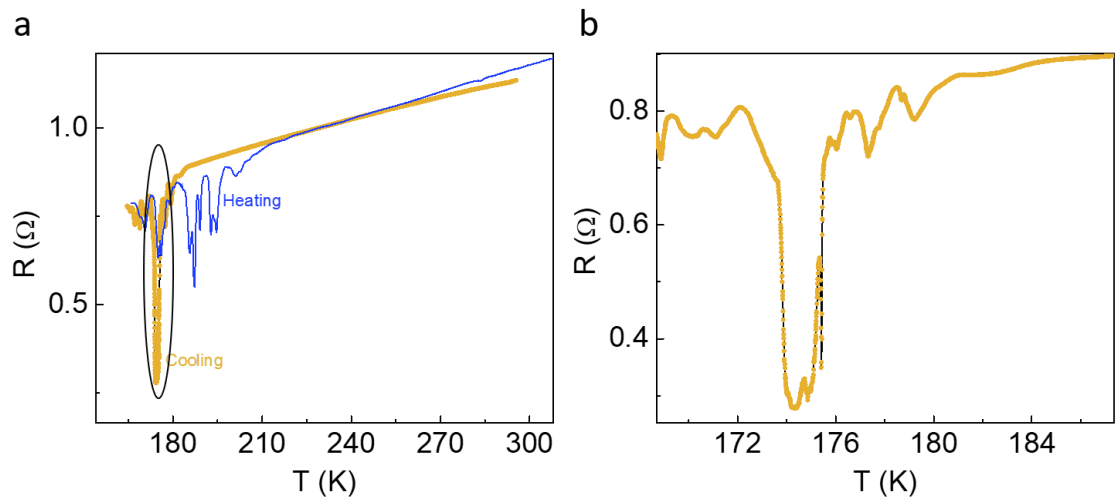

**Figure S19. Incomplete transition in sample P20918FE0_02.** (a) Resistance vs temperature for both cooling and heating cycles, showing incomplete transition. The resistance drops sharply around $T = 175$ K and recovers with an oscillatory behaviour around the drop as shown in the magnified scale in (b) for the cooling cycle.

P21018FE0_04 also demonstrated 5 % drop in resistance at $T \sim 200$ K, which was repeated over several thermal cycles, as shown in Figure S20. This suggests the possibility of only a small area of the sample transitioning from a normal to superconducting state.

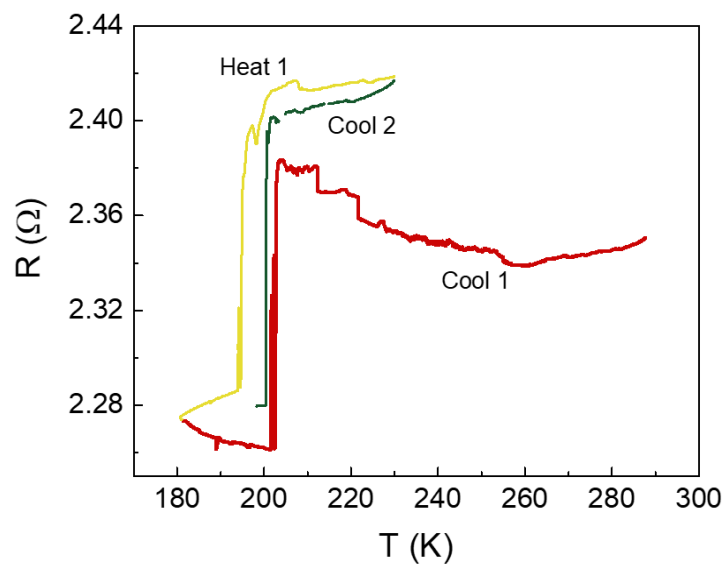

**Figure S20. Incomplete transition in sample P21018FE0_04.** Resistance vs temperature for several thermal cycles, showing a repeatable drop (rise) during cooling (heating) cycles.

We have also observed multiple resistance drops, presumably due to local superconducting transitions arising from inhomogeneous distribution of nanoclusters (or other extraneous effects). This is illustrated with sample P20219FE0_18 in Figure S21.

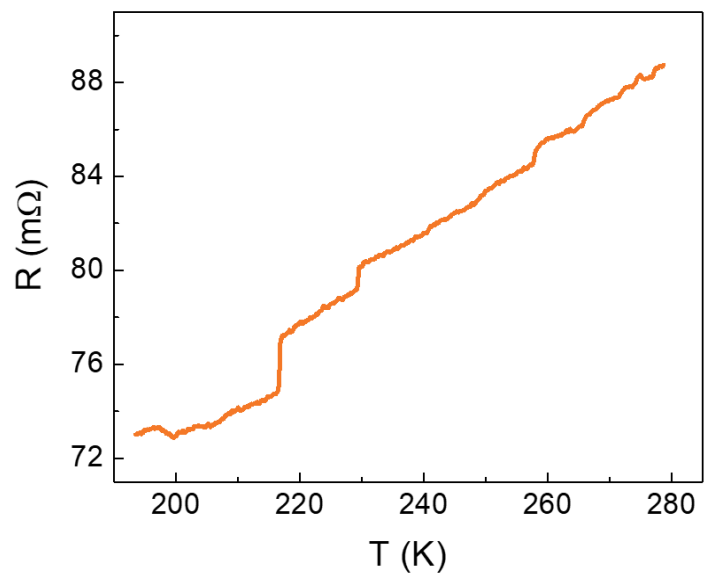

**Figure S21. Incomplete multiple transitions in sample P20219FE0_18.** Resistance vs temperature showing repeated drops at several temperatures.

**Section S15. Re-entrant transitions in resistance**

Several of the samples, pristine or aged, showed a surprising "re-entrant behaviour", with the resistance fluctuating between superconducting state (zero-resistance state) and the normal state at various temperature. This effect is illustrated with Sample P20119FE0_13 where the re-entrant behaviour can be seen at $T = $ 175 K, 150 K, 100 K (in this particular cycle). This "re-entrant" behaviour was observed in several cycles, as shown in Figure S22, indicating the unstable nature of the transition in this sample. Importantly, we did not observe any signature of any of the transitions in the mutual inductance, which could be explained if the superconducting region is far away from the coils.

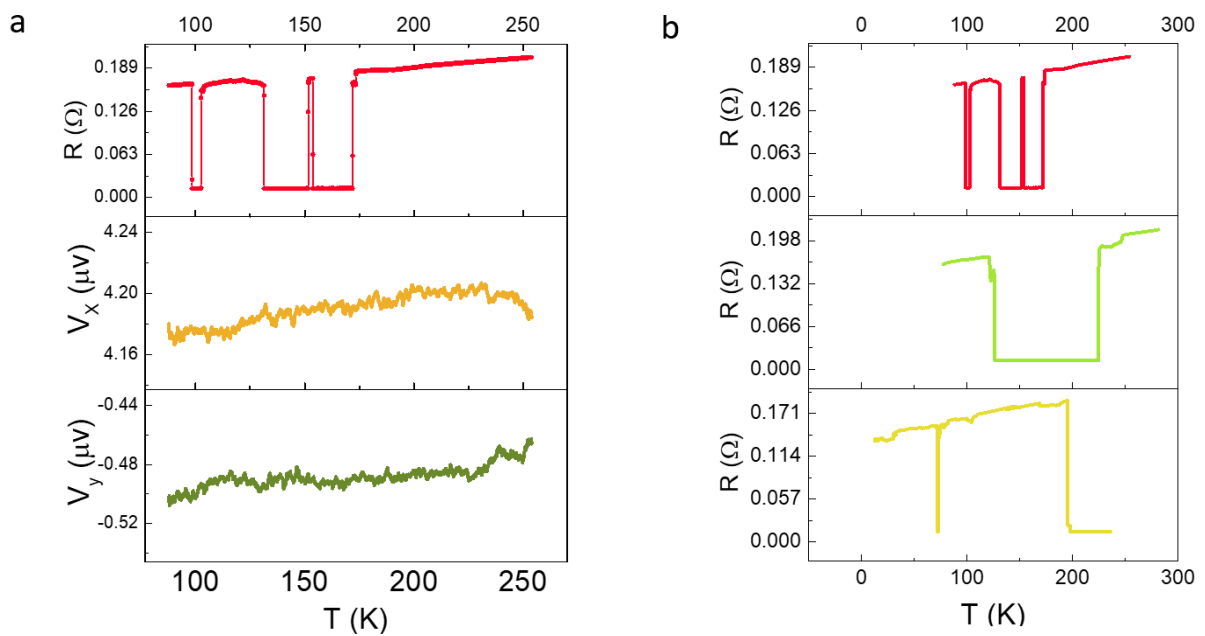

**Figure S22. Re-entrant transition in resistance.** (a) Resistance vs. temperature showing repeated transitions between zero and finite resistance states. (b) Re-entrant behaviour in resistance observed in three different thermal cycles.

**Section S16. Random telegraphic signal in resistance**

Some of the samples have displayed two-state random telegraphic fluctuations in resistance within over some specific temperature range. Such examples are given below.

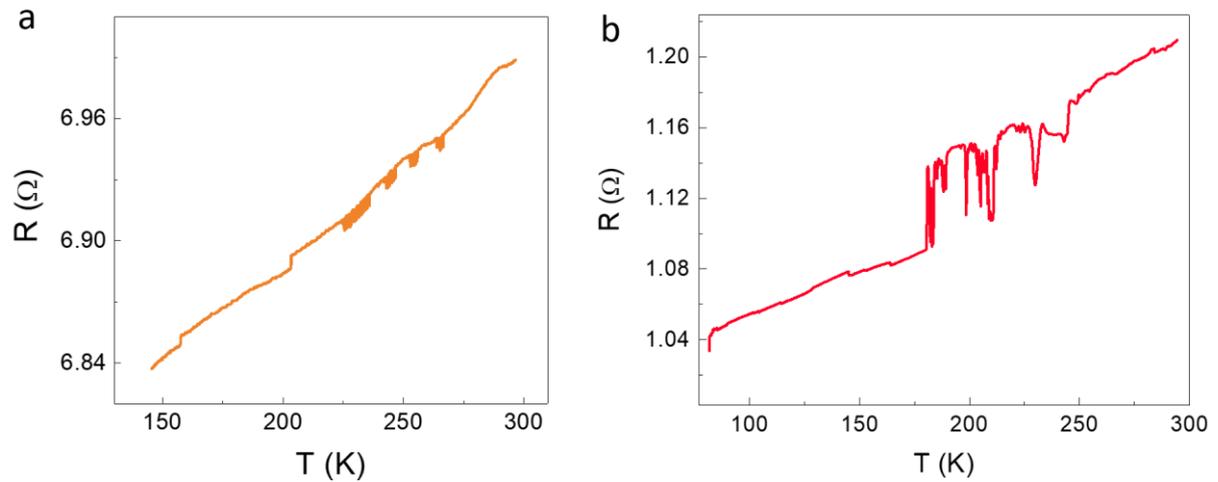

**Figure S23. Two state fluctuations in resistance.** (a) Two-state fluctuations in resistance in sample P21018FE0_14 around $T = 240 \pm 20$ K. (b) Two-state fluctuations in resistance in sample P21118FE0_10 around $T = 205 \pm 25$ K

## Section S17. Examples of simultaneous change in $R$, $V_x$ and $V_y$

The details of the samples that have shown simultaneous features in resistance, and inductive response are shown in Figure S24 below. While the transition in sample P21218FE0_12, as shown in Figure S24, are incomplete, the data alludes to small regions of the sample becoming superconducting at different temperatures, manifested in simultaneous changes in both $R$ and $V_x$. $\Delta R$ has been calculated by subtracting a (smoothly varying background) resistance from the sample.

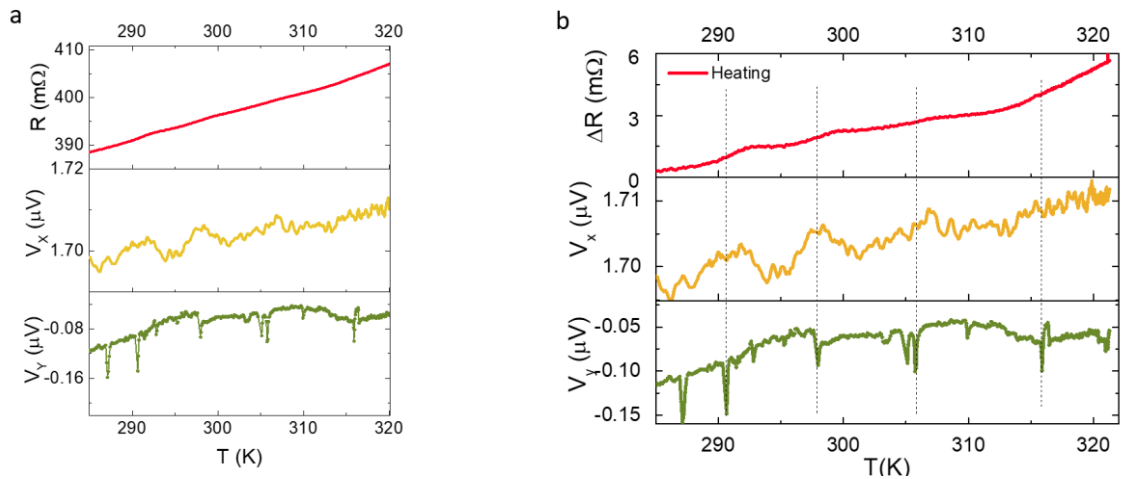

**Figure S24. Simultaneous change in resistance and inductive response.** (a) $R$, $V_x$ and $V_y$ for P21218FE0_12. (b) $\Delta R$, $V_x$ and $V_y$ for the same sample (The pick-up voltage $V_x$ has not been background corrected).

Simultaneous changes in resistance and inductive response was also observed in sample P20119FES_07 around $T = 263$ K and presented in Figure S25 below.

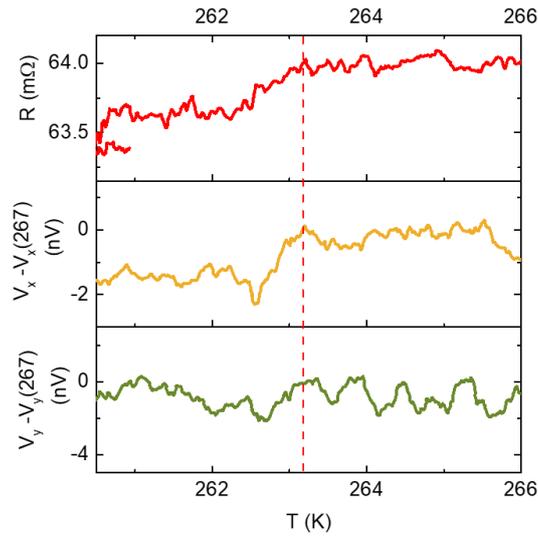

**Figure S25. Simultaneous changes in resistance and inductive response in P20119FES_07.** Temperature-dependence of $R$, $V_x$ and $V_y$ for P20119FES_07.

Sample P21218FE0_08 also exhibited simultaneous features (at $T = 288$ K) in resistance and inductive response. $\Delta R$ has been calculated by subtracting a (background) resistance by fitting the red dashed line as shown in Figure S26a.

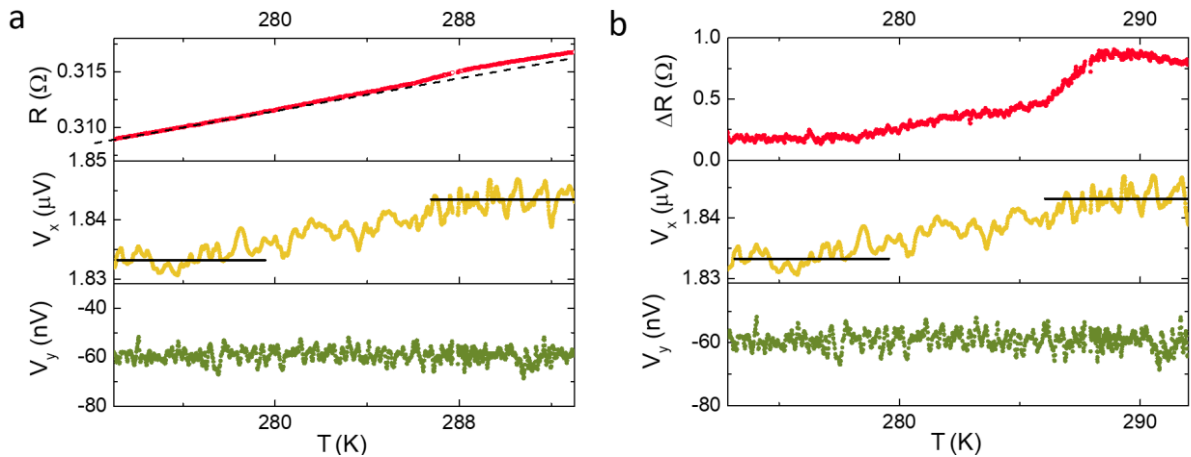

**Figure S26. Simultaneous changes in resistance and inductive response in P21218FE0_08.** (a) $R$, $V_x$ and $V_y$ for P21218FE0_08. The drop in $R$ at $T = 288$ K is accompanied by a drop in $V_x$ (not corrected for background). (b) $\Delta R$, $V_x$ and $V_y$ for the same sample.

Sample P20219FE0_18 also showed simultaneous changes in these parameters around $T = 160$ K (Figure S27)

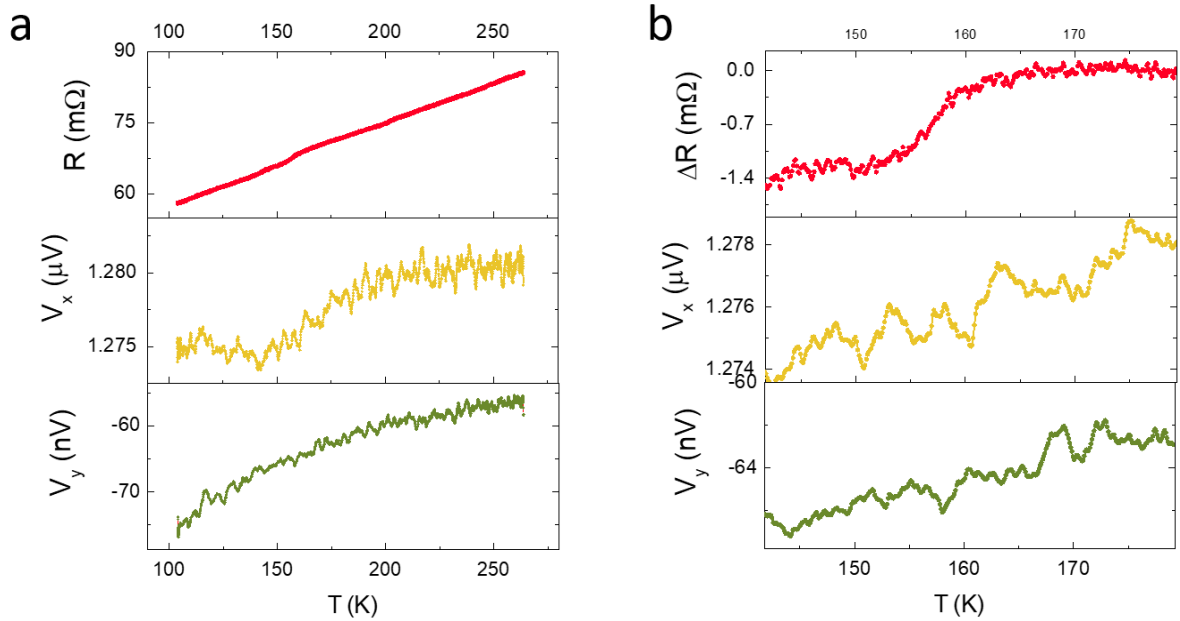

**Figure S27. Simultaneous changes in resistance and inductive response in P20219FE0_18.** (a) $R$, $V_x$ and $V_y$ for sample P20219FE0_18. The drop in $R$ at $T = 160$ K coincides with a drop in $V_x$. (b) $\Delta R$, $V_x$ and $V_y$ for the same sample.

In Sample P20119FE0_08 a simultaneous change in resistance and inductive response could be observed at $T = 261$ K (Figure S28).

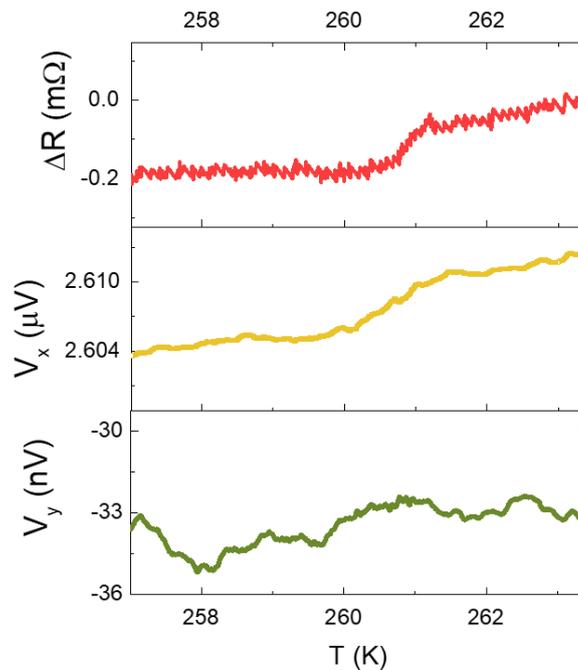

**Figure S28. Simultaneous change in resistance and inductive response in P20119FE0_08.** Temperature dependence of $\Delta R$, $V_x$ and $V_y$ for sample P20119FE0_08. The drop in $R$ at $T = 261$ K is accompanied by a drop in $V_x$ (not background corrected).

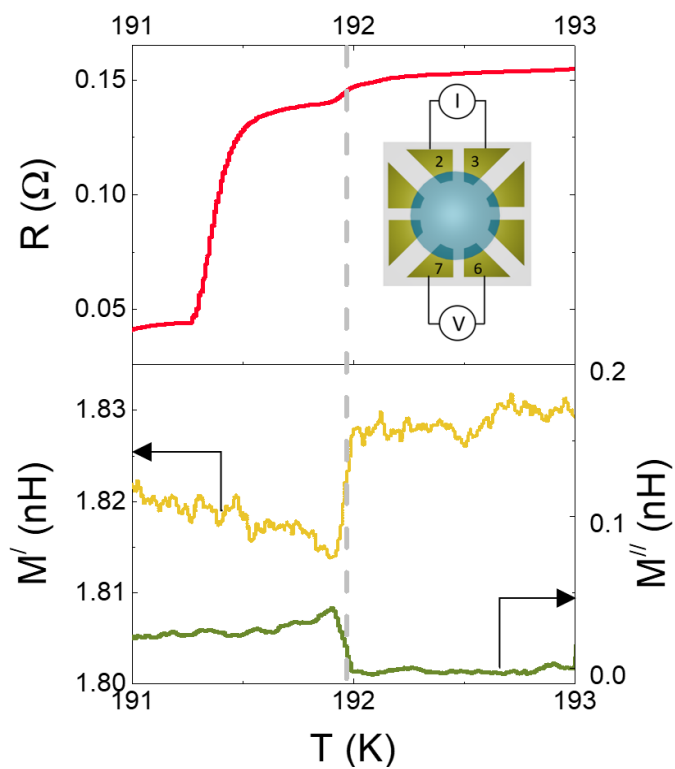

**Figure S29. Simultaneous change in resistance and inductive response in P20319FEE_05.** Simultaneous measurement of resistance (top) and inductive response (bottom) in Device P20319FEE_05 (Table S1, SI) as a function of temperature. The inset in the upper panel schematically shows the lead configuration used for electrical measurements. $M'$ and $M''$ are real and imaginary components of the mutual inductance between the drive and sense coils.

**Section S18. Evidence of inductive transition without resistive features**

Some of the samples have shown a strong inductive response as the temperature has been reduced. The response was observed only in susceptibility measurements, with no simultaneous signature in resistance. In Sample P21218FE0_05, the transition temperature for such a response was at $T = 201$ K, as shown in Figure S9a. In sample P20319FEE_18, this was observed at $T = 182$ K.

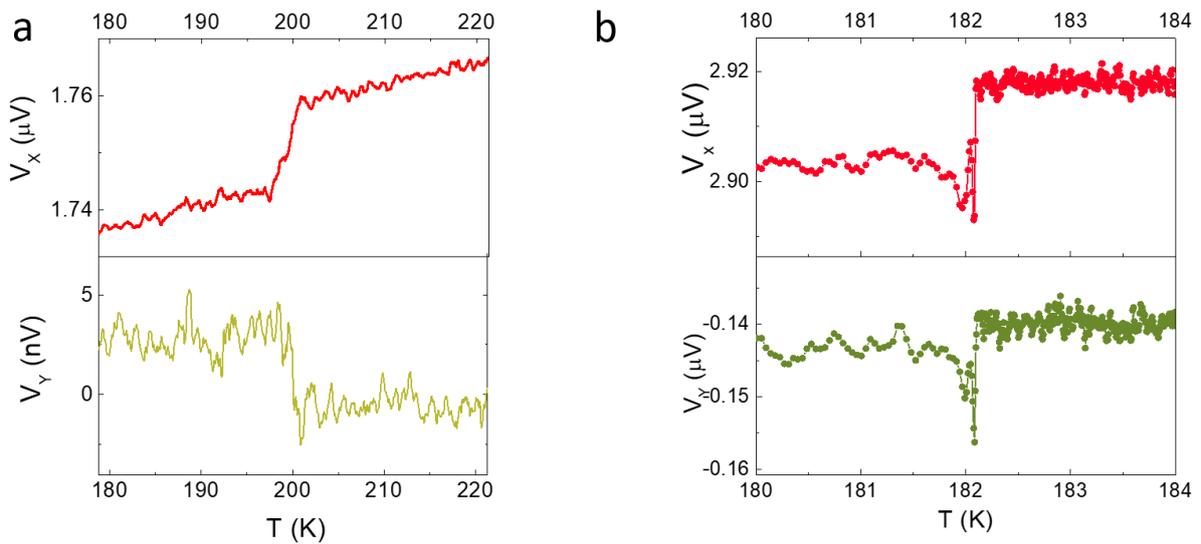

**Figure S30. Signatures of transition in inductive response** in (a) P21218FE0_05, and (b) P20319FEE_18.

## Section S19. Halperin-Nelson Fit

The resistive transition in several devices could be fitted with the Nelson-Halperin formulation that is applicable to the Berezinskii-Kosterlitz-Thouless (BKT) transition in thin superconducting films, and is given as (4)

$$R = \frac{R_N}{1+\left(A \sinh\frac{B}{\sqrt{T-T_c}}\right)^2} \quad (S6)$$

Here $R_N$ and $T_c$ are the normal state resistance and BKT transition temperature, respectively. $A$ and $B$ are numerical terms that depend on superfluid density and other parameters (5, 6).

Intriguingly, although the sample thickness (~ 25 – 100 nm) is not necessarily smaller than $\xi$, we find that the Nelson-Halperin formula, which is applicable to the BKT transition in thin superconducting film, does seem to fit the $T$-dependence of $R$. We however emphasize that the BKT framework may not be unique in describing the transition, and further understanding of the quantum state of the nanostructure at low $T$ is required for a more quantitative analysis.

Figure S31 presents the fitting in three different samples, two of which show complete transition to the zero-resistance state below the transition temperature. Intriguingly, Equation S6 could describe the transition in aged devices as well, as illustrated with P20319FEE_06 in Figure S32.

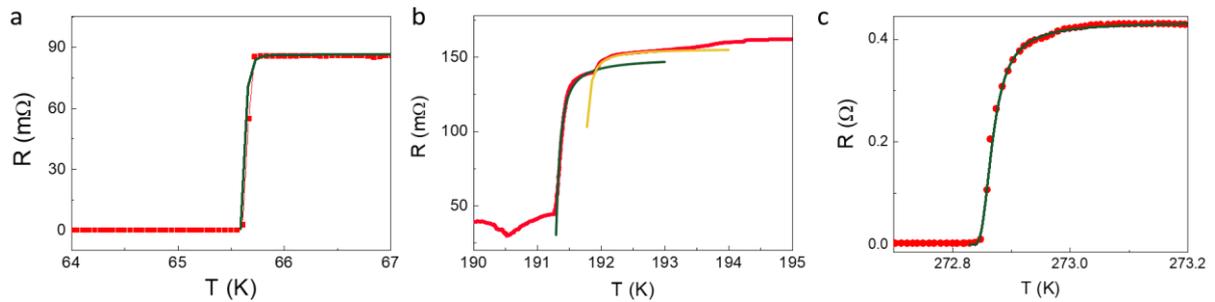

**Figure S31. Fitting of Halperin-Nelson formula for pristine samples:** (a) The solid line shows the Halperin-Nelson fit to the data (red points), which captures the variation of R with T reasonably well for P20319FEE_03. (b) H-N fit for P20319FEE_05. (c) H-N fit for P20319FEE_06. The fitting parameters are given in Table S2.

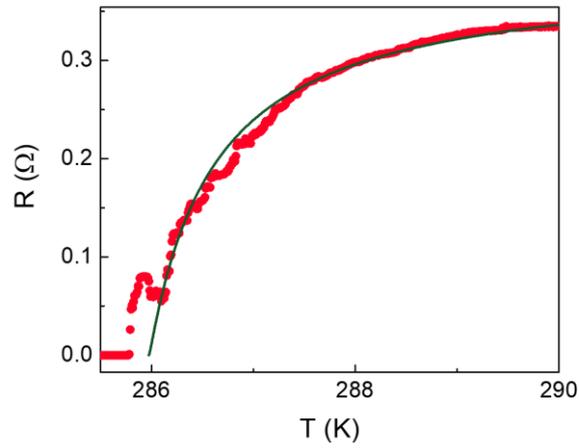

**Figure S32. Fitting of Halperin-Nelson formula for aged sample:** The solid line shows the Halperin-Nelson fit to the transition data (red points) for P20319FEE_06 two weeks after it was prepared (and subsequently subjected to electrical bias/thermal cycles).

**Table S2: Details of Halperin-Nelson fitting parameters**

| Sample No | $T_C$ | $R_N$ (Ω) | A | B |
|---|---|---|---|---|
| P20319FEE_03 | 65.55 | 0.086 | 0.01 | 1.5 |
| P20319FEE_05 | 191.25 | 0.15 | 0.38 | 0.5 |
| P20319FEE_05 | 191.7 | 0.0156 | 0.25 | 0.5 |
| P20319FEE_06 | 272.83 | 0.463 | 0.1 | 0.6 |
| P20319FEE_06 | 285.97 | 0.39 | 4.0 | 0.2 |

**Section S20. Volume susceptibility in lead**

Lead Shot 99.995 % trace metal basis Lot No. MKBS8698V was purchased from Sigma Aldrich. The chemical was used as received without further processing. The data in Figure S33 below corresponds to a 90 mg pellet retrieved from this stock.

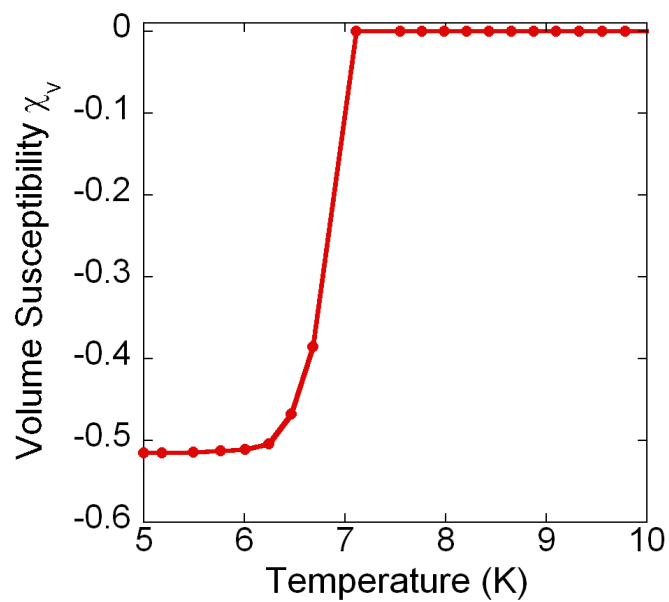

**Figure S33. Volume susceptibility of lead pellet**

## Section S21. Noise characteristics in magnetic measurements

Two different varieties of unusual "noise" patterns were observed in magnetic measurements. Both classes of patterns occur significantly above the instrument noise threshold, and therefore suggest a possible physical origin related to the sample as opposed to instrument artefacts. We note that most Au-Ag NS samples that undergo transitions within the squid measurement system show unusually greater noise as compared to the normal instrument threshold. We could not locate any measurement protocol anomalies/data acquisition problems/processing errors that could possibly give rise to these patterns. Indeed, enhanced noise below the transition was not observed in standard Pb pellets that were purchased from Sigma Aldrich (Lead Shot 99.995 % trace metal basis Lot No. MKBS8698V) that were measured using similar protocols. Based on our present understanding, the two categories of repeat patterns that are described below may or may not have a physical origin and further study is needed to understand these more completely.

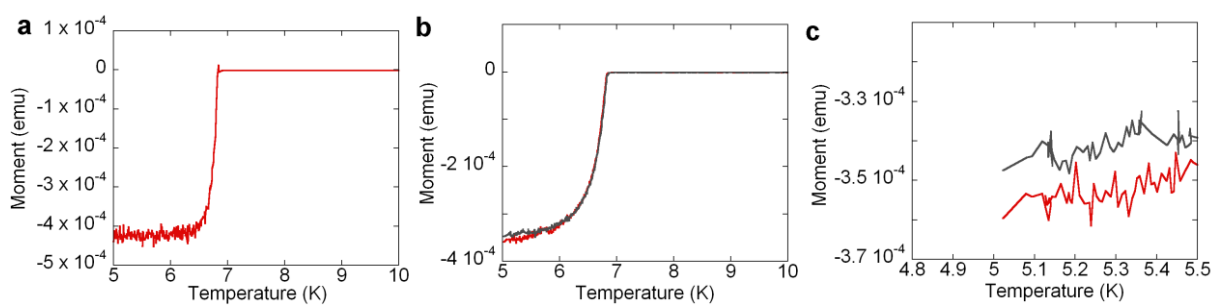

**Figure S34. Magnetic Properties of Pb-CTAB Pellets**. **(a)** ZFC scan of pellet 1 described in text below **(b)** Concurrent ZFC scans on pellet 2. **(c)** Expanded view of 5-5.5 K region for pellet 2.

The data presented in Figure 3c show a peculiar pattern that recurs over three scans (0.01 T, 0.1 T and 1 T) that are taken as a function of temperature. This pattern is significantly greater (by about 4 orders) in its absolute amplitude as compared to instrument noise, and further occurs only below the transition temperature. Similar pattern repetition is observed in other experiments involving multiple scans over a single pellet. Since this pattern repeats with different scans but not with a complete reproducibility, we hypothesized it to be related to thermo-mechanical properties of pellets involving a superconductor-surfactant combination. To check this hypothesis, we prepared pellets of a more studied conventional superconductor (lead) with CTAB. Figure S34 shows the properties of two different types of lead-CTAB pellet. Here pellet 1 has been prepared by pressing a mixture containing 75 mg of lead powder (purchased from Sigma Aldrich (325 Mesh, >= 99 % Trace Metals Basis Lot No. STBG5744V) and 25 mg of CTAB (Sigma Aldrich >=98 % Lot No. SLBT7256) in a titanium die. On the other hand pellet 2 was made by solution processing. Equal volumes of lead powder (grain size 37 micron) and CTAB were taken and dissolved in 0.2 ml of methanol with centrifugation. The methanol was left to dry and again 0.2 ml of methanol was added to the solid mass. This process of dissolving and drying methanol was done 5 times. The final dried product was then pressed in a titanium die to form the pellet. As evident in these studies, additional noise-like patterns do indeed emerge in both cases which are higher than the noise threshold of the instrument. It is therefore evident that CTAB-superconductor composites exhibit unusual noise below the transition temperature unlike pure superconductor pellets (see Figure S33 for a pure lead pellet). We were however unable to achieve a precise replication of the noise pattern to the degree observed in Figure 3c. It is hypothesized that this is due to an intrinsically greater mechanical plasticity of the lead-CTAB pellets currently prepared by us. Further optimization of lead-CTAB pellet making protocols may enable the design of pellets with susceptibility patterns that repeat across scans in these systems.

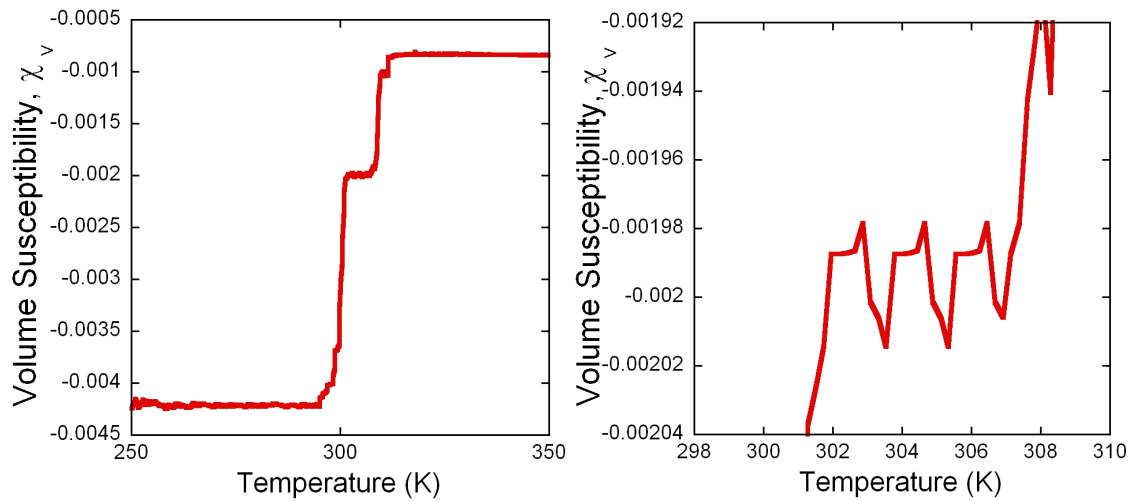

**Figure S35**. ZFC scan of pellet P10218PE0_01 showing repeat patterns in susceptibility over $302 - 307$ K.

In addition to the above patterns, a second, more significant variety of repetition was also observed in certain cases. This corresponds to repetition of certain regions of the susceptibility curve at various temperatures as exemplified in Figure S35. This feature is again exclusively observed in certain samples that appear to undergo transitions at measurably low temperatures. Such repeat patterns are absent in non-superconducting samples, conventional superconductors as well as apparently superconducting samples with transition temperatures that are too high to be measured. While we cannot rule out the potential of measurement artefacts arising from the instrument, the absence of similar features in other classes of materials described above does suggest a sample related origin to these features as well. We hypothesize that these patterns could possibly arise from persistent currents generated within the samples during the occurrence of a transition or else could simply be a yet unknown measurement artefact. Unfortunately, our present understanding of the material does not allow us to resolve this matter.

## Section S22. Transition in resistance for different Au and Ag mole fractions

Figure S36 shows the resistance of three different samples as a function of temperature. All three samples are derived from the same initial stock of NS. Subsequently varying thicknesses of Au have been grown over each using the procedures described in Section S1.

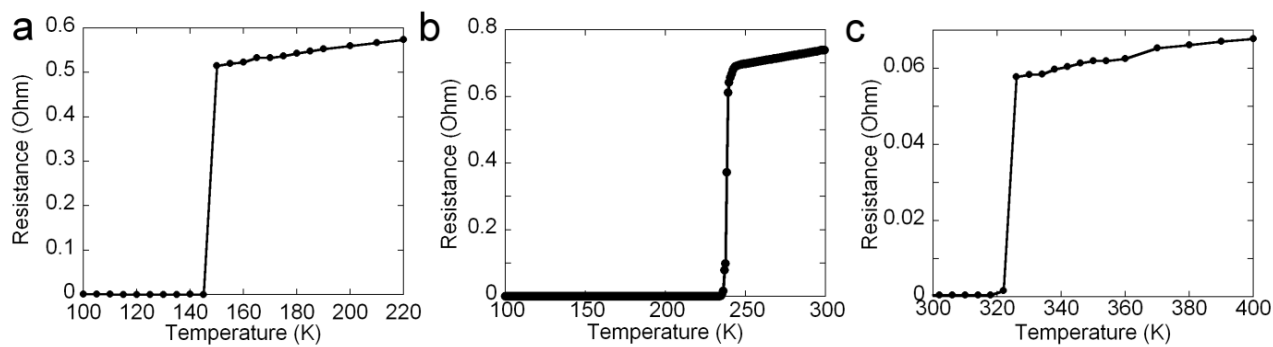

**Figure S36.** Transitions in resistance as observed for three different NS samples. (a) $x_{Au} = 0.87$ (b) $x_{Au} = 0.83$ (c) $x_{Au} = 0.80$.

**Movie S1:**

The movie can be viewed on the following link:

[LINK TO MOVIE S1](LINK TO MOVIE S1)

Definitions:

- $P_{O_2}$: Oxygen level in glove box in ppm
- $R_{NS}$: Normal state resistance at room temperature
- TCR: Temperature co-efficient of resistance, defined as $\text{TCR} = \frac{1}{R_{273}} \frac{dR}{dT}$, with $R_{273}$ is the resistance of the sample at $T = 273$ K unless otherwise mentioned
- $R_{\text{4P\_MAX}}$: Maximum four probe resistance in the sample at room temperature
- $R_{\text{4P\_MIN}}$: Minimum four probe resistance in the sample at room temperature
- Samples are named as follows: Protocol (P1/P2)Date(MMYY)Pellet or Film (P/F)Encapsulation (E0: None/EW: Wax/EC: Copper/ES: Silver/EE: Epoxy/EN: Nail Enamel)_SN (01, 02 etc).

For example, P11216PE0_02 implies

P1: Protocol 1
1216: December 2016
P: Pellet
E0: No Encapsulation
02: Serial Number 2 for the month of December 2016.

### Table S1: Details of measured samples

| Sl No. | Sample Code | $P_{O_2}$ (ppm) | $R_{NS}$ ($\Omega$) | Sample Nature/TCR | General Remarks |
|---|---|---|---|---|---|
| 1 | P11116FE0_01 | Prepared in Air | >1 GOhm, 2 Probe only | • Insulator | • Poor contacts and cross linking <br> • $R$ measurement in ambient |
| 2 | P11116PE0_02 | Prepared in Air | NA | | • MH Sweep at 100 K <br> • Paramagnetic with possible weak ferromagnetism |
| 3 | P11116PE0_03 | <5 | NA | | • Weak diamagnetism from 5 K to 300 K at 0.01 T <br> • Sample becomes more paramagnetic at low $T$ |
| 4 | P11216PE0_01 | <5 | NA | | • Weak diamagnetism from 5 K to 400 K at 0.01 T <br> • Sample becomes less diamagnetic at low $T$ (below 50 K) <br> • High noise as well as small sharp rise in paramagnetism at 370 K |
| 5 | P11216PE0_02 | <5 | NA | | • Weakly diamagnetic Sample <br> • Becomes more diamagnetic at low $T$ <br> • Small Sharp drop in susceptibility at 60 K <br> • Slower unresolved (6 %) drop at $5-8$ K |
| 6 | P10117PE0_01 | <5 | NA | | • Weakly Diamagnetic Sample |
|  |  |  |  |  | • Becomes more diamagnetic at $T < 331$ K and $T > 331$ K |
|  |  |  |  |  | • Becomes least diamagnetic at 331 K |
|  |  |  |  |  | • Multiple sharp drops in susceptibility at 66, 114, 185, 241, 253 K |
|  |  |  |  |  | • Slower unresolved drop at $5 - 8$ K |
|  |  |  |  |  | • Sharp increases in susceptibility at 260 and 338 K |
| 7 | P10117PE0_02 | < 5 | NA |  | • Weakly diamagnetic sample. <br> • 15 % drop in susceptibility at $125 - 140$ K <br> • 10 % rise between $120 - 50$ K <br> • 10 % drop between $5 - 50$ K. |
| 8 | P10217PE0_01 | <5 | NA |  | • Weakly diamagnetic sample. <br> • 5 % drop in susceptibility at 120 K <br> • Sample becomes less diamagnetic as $T$ falls below 10 K. |
| 9 | P10217PE0_02 | < 5 | NA | NA | • Paramagnetic |
| 10 | P10217PE0_03 | < 5 | NA |  | • Weakly diamagnetic sample <br> • Reentrant transitions at $160 - 180$ K <br> • Small drop at 123 K <br> • Sample becomes less diamagnetic below 50 K |
| 11 | P10317PE0_01 | < 5 | NA |  | • Weakly diamagnetic Sample (3 T Data) |
| 12 | P10317PE0_02 | < 5 | NA |  | • Weakly diamagnetic, with a dip (16 %) starting at 20 K and a small rise at 8 K. |
| 13 | P10517PE0_01 | < 5 | NA |  | • Weakly diamagnetic sample <br> • Imperfectly cleaned |
| | | | | | • Small (7 %) Fall in susceptibility at 121.5 K (0.01 T)<br>• Shifts slightly to lower temperature (121 K) at higher field (0.015 T)<br>• Indistinct data due to large background |
| 14 | P10617PE0_01 | < 5 | NA | | • Weakly diamagnetic<br>• Sharp increase in susceptibility at 250 K (Similar to re-entrant transitions in resistivity) |
| 15 | P10717FE0_01 | <5 | NA | NA | • Contact broke during measurement |
| 16 | P10917FE0_01 | < 5 | 2 | • Metallic<br>• $3.9 \times 10^{-3}$ | |
| 17 | P10917PE0_02 | < 5 | NA | | • Weakly diamagnetic/paramagnetic<br>• Paramagnetic 174 − 371 K<br>• Diamagnetic at other temperatures. |
| 18 | P10917FE0_03 | < 5 | 0.55 | $5 \times 10^{-4}$ | • Transition at 150 K<br>• Observation of zero resistance |
| 19 | P11017FES_01 | < 5 | NA | $9 \times 10^{-3}$ | • Resistance $1 \times 10^{-3}$ at 300 K<br>• 25 nm Silver deposited over sample through thermal evaporation<br>• Consistent with superconducting state from 150 − 350 K<br>• Observation of zero resistance |
| 20 | P11017FE0_02 | <5 | 1.79 | $1 \times 10^{-3}$ | • Field dependent resistance (see main text Figure 3) with transitions<br>• Observation of zero resistance |
| 21 | P11017FE0_03 | < 5 | 11.0 | $6.4 \times 10^{-4}$ | • Unusual double transition at 180 K (99.8%) and 200 K (0.2 %) |
| 22 | P11017FE0_04 | < 5 | 4.3 | $2 \times 10^{-3}$ | • metallic |
| 23 | P11017FE0_05 | < 5 | $4.5 \times 10^{-2}$ (at 310 K) | NA | • Incomplete transition at 300 K<br>• 67% |

| 24 | P11017FE0_06 | < 5 | NA | $-2 \times 10^{-4}$ | • Resistance $3 \times 10^{-3}$ at 300 K<br><br>• Consistent with superconducting state from 100-400 K<br>• Observation of zero resistance |
|---|---|---|---|---|---|
| 25 | P11017FE0_07 | < 5 | 0.42 | $1 \times 10^{-3}$ | • Transition at 219 K<br>• Observation of zero resistance |
| 26 | P11017FES_08 | < 5 | NA | $7 \times 10^{-4}$ | • Film with 180 nm silver deposition on top of sample |
| 27 | P1017FES_09 | < 5 | NA | $8 \times 10^{-4}$ | • Film with 200 nm silver deposition on top of sample |
| 28 | P11117PE0_01 | <5 | NA | NA | • Transitions at Field dependent temperatures. See Figure 3 of main text. |
| 29 | P11117PE0_02 | < 5 | NA |  | • Noisy data<br>• Weakly diamagnetic |
| 30 | P11117PE0_03 | <5 | NA | NA | • Strong diamagnetism (−0.037) over 150 K - 350 K |
| 31 | P11217PE0_01 | <5 | NA | NA | • FC/ZFC Scan with sample transition at 232 K |
| 32 | P10118PE0_03 | <5 | NA | NA | • Strong diamagnetism at room temperature |
| 33 | P10218PE0_01 | <5 | NA | NA | • Significantly diamagnetic (−0.00083) at 350 K.<br>• More diamagnetic when cooled with three different transitions at 312 K, 309 K and 301 K.<br>• Susceptibility at 290 K is −0.004. |
| 34 | P20918FE0_01 | 10-20 | 24-30 | • Mixed<br>• $9.81 \times 10^{-4}$ | • Metal-insulator transition in different thermal cycles<br>• See Figure S11 |

| 35 | P20918FE0_02 | 10-20 | 1.0 | - Metallic<br>- $2 \times 10^{-3}$ | - Incomplete transition<br>- Abrupt jump in $R$ at $T \sim 176$ K<br>- Repeated in multiple thermal cycles<br>- See Figure S19 |
|---|---|---|---|---|---|
| 36 | P20918FE0_03 | 10-20 | 0.57-0.63 | - Metallic<br>- $2.92 \times 10^{-3}$ | - Small abrupt jumps in the resistance in successive thermal cycles<br>- Hysteretic |
| 37 | P21018FE0_01 | 10-20 | 0.5 | - Metallic<br>- $4.05 \times 10^{-4}$ | |
| 38 | P21018FE0_02 | 10-20 | 0.006 | - Metallic<br>- $4.3 \times 10^{-3}$ | - Small jumps observed<br>- V-I flat<br>- Observation of zero resistance |
| 39 | P21018FE0_03 | 10-20 | 0.075 | - Metallic<br>- $2.24 \times 10^{-3}$ | |
| 40 | P21018FE0_04 | 10-20 | 2.352 | - Mixed above<br>- Tc $\sim 210$ K<br>- $6.06 \times 10^{-5}$ | - Incomplete transition in $R$<br>- See Figure S20 |

| 41 | P21018FE0_05 | 10-20 | 0.585 | - Metallic<br>- $2.43\times10^{-4}$ | - Non-repeatable, incomplete transition |
| 42 | P21018FE0_06 | 10-20 | 0.610 | - Mixed<br>- $6.46\times10^{-4}$ | - Metal-insulator transition in different thermal cycles<br>- Hysteretic nature of $R$ vs $T$ |
| 43 | P21018FE0_07 | 10-20 | 0.297 | - Metallic<br>- $8.37\times10^{-4}$ | - Two state fluctuations in $R$ |
| 44 | P21018FE0_08 | 10-20 | 1.13 | - Metallic<br>- $1.47\times10^{-4}$ | - Abrupt, small changes in $R$ ($\Delta R \sim 0.45\ \%$) around $T\ =\ 159.5\ \text{K}$ |
| 45 | P21018FE0_09 | 10-20 | 0.023 | - Insulating<br>- $-3.11\times10^{-3}$ | |
| 46 | P21018FE0_10 | 10-20 | 0.036 | - Metallic<br>- $2.14\times10^{-3}$ | |

| 47 | P21018FE0_11 | 10-20 | NA | NA | No electrical contact<br>No transition in inductive response |
|---|---|---|---|---|---|
| 48 | P21018FE0_12 | 10-20 | 0.185 | Metallic<br>$2.31 \times 10^{-3}$ | |
| 49 | P21018FE0_13 | 10-20 | 0.569 | Metallic<br>$9.7 \times 10^{-3}$ | Metallic |
| 50 | P21018FE0_14 | 10-20 | 6.96 | Metallic<br>$1.78 \times 10^{-4}$ | Two state fluctuations in $R$ around $T = 240 \pm 20$ K<br>See Figure S23 |
| 51 | P21018FE0_15 | 10-20 | 1.18 | Metallic<br>$2.43 \times 10^{-4}$ | |
| 52 | P21018FE0_16 | 10-20 | 1.35 | Metallic<br>$7.72 \times 10^{-4}$ | |
| 53 | P21118FE0_01 | 10-20 | 1.0 | - Metallic<br>- $5.71 \times 10^{-4}$ | |
| 54 | P21118FE0_02 | 10-20 | 4.6 | - Metallic<br>- $4.2 \times 10^{-4}$ | |
| 55 | P21118FE0_03 | 10-20 | 0.0952 | - Metallic<br>- $1.68 \times 10^{-3}$ | |
| 56 | P21118FE0_04 | 10-20 | 6.65 | - Metallic<br>- $5.66 \times 10^{-4}$ | - Hysteresis in heating and cooling<br>- Small ($\Delta R \sim 0.3\,\%$) drops in $R$ around $T = 275\,K$ |
| 57 | P21118FE0_05 | 10-20 | 4.69 | - Metallic<br>- $1.61 \times 10^{-3}$ | |
| 58 | P21118FE0_06 | 10-20 | 6.25 | - Metallic<br>- $8.78 \times 10^{-4}$ | |

| 59 | P21118FE0_07 | 10-20 | 7.2 | - Mixed<br>- $4.31 \times 10^{-4}$ | - Metal-insulator transition in different thermal cycles |
| --- | --- | --- | --- | --- | --- |
| 60 | P21118FE0_08 | 10-20 | 0.265 | - Metallic<br>- $4.01 \times 10^{-4}$ | |
| 61 | P21118FE0_09 | 10-20 | 1.6 | - Metallic<br>- $4.13 \times 10^{-4}$ | |
| 62 | P21118FE0_10 | 10-20 | 1.2 | - Metallic<br>- $5.68 \times 10^{-4}$ | - Two state fluctuations in $R$ around $T = 205 \pm 25$ K<br>- See Figure S23 |
| 63 | P21118FE0_11 | 10-20 | 0.305 | - Metallic<br>- $1.21 \times 10^{-3}$ | - Repeatable, abrupt, small jumps ($\Delta R \sim 4\%$) observed |
| 64 | P21118FE0_12 | 10-20 | 0.74 | - Metallic<br>- $8.63 \times 10^{-4}$ | - TCR at calculated $T = 269$ K |
| 65 | P21118FE0_13 | 10-20 | 2.0 | • Mixed<br>• $3.72 \times 10^{-4}$ | • Abrupt changes in $R$<br>• Metal to insulator transition<br>• Measured down to $T = 4.2$ K |
| 66 | P21218FE0_01 | 10-20 | 2.6 | • Mixed<br>• $7.78 \times 10^{-4}$ | |
| 67 | P21218FE0_02 | 20-30 | 0.1323 | • Metallic<br>• $1.34 \times 10^{-3}$ | |
| 68 | P21218FE0_03 | 20-30 | 0.105 | • Metallic<br>• $1.25 \times 10^{-3}$ | • Abrupt changes in $R$<br>• No simultaneous transition in inductive response |
| 69 | P21218FE0_04 | 30-40 | NA | • NA | • No electrical contacts<br>• No change in susceptibility |
| 70 | P21218FE0_05 | 30-40 | NA | • NA | • Change in inductive response observed at $T = 201$ K<br>• No electrical contact<br>• See Figure S30 |
| 71 | P21218FE0_06 | >50 | NA | - NA | - No electrical contacts<br>- No change in susceptibility |
| 72 | P21218FE0_07 | >50 | 5.3 | - Metallic<br>- $3.74 \times 10^{-4}$ | - TCR calculated at $T = 280$ K |
| 73 | P21218FE0_08 | >50 | 0.315 | - Metallic<br>- $1.36 \times 10^{-3}$ | - Possible signature of simultaneous change in $R$ and inductive response<br>- Broad transition<br>- See Figure S26 |
| 74 | P21218FE0_09 | >50 | 0.395 | - Metallic<br>- $1.16 \times 10^{-3}$ | |
| 75 | P21218FEW_10 | >50 | 3.4 | - Insulating<br>- $-3.14 \times 10^{-3}$ | - Abrupt changes in $R$<br>- No simultaneous change in inductive response |
| 76 | P21218FE0_11 | >50 | 1.25 | - Insulating<br>- $1.002 \times 10^{-2}$ | |

| 77 | P21218FE0_12 | >50 | 0.400 | • Metallic<br>• 1.34x10$^{-3}$ | • Incomplete transitions at $T = 300 \pm 10$ K<br>• Simultaneous signature in $R$ and inductive response<br>• See Figure S24 |
|---|---|---|---|---|---|
| 78 | P21218FE0_13 | >50 | 0.300 | • Metallic<br>• 1.36 x10$^{-3}$ | • $R_{\text{4P\_MAX}}$: 0.334 Ω<br>• $R_{\text{4P\_MIN}}$: 0.0748 Ω |
| 79 | P20119FE0_01 | >50 | 0.0498 | • Metallic<br>• 6.07x10$^{-4}$ | • $R_{\text{4P\_MAX}}$: 0.17289 Ω<br>• $R_{\text{4P\_MIN}}$: 0.00503 Ω |
| 80 | P20119FE0_02 | >50 | 0.0543 | • Metallic<br>• 9.66x10$^{-4}$ | • $R_{\text{4P\_MAX}}$: 0.13454 Ω<br>• $R_{\text{4P\_MIN}}$: 0.05604 Ω |
| 81 | P20119FE0_03 | >50 | 0.310 | • Metallic<br>• 3.56x10$^{-4}$ | • $R_{\text{4P\_MAX}}$: 0.57319 Ω<br>• $R_{\text{4P\_MIN}}$: 0.31334 Ω |
| 82 | P20119FE0_04 | >50 | 0.170 | • Metallic<br>• 1.18x10$^{-4}$ | |

| 83 | P20119FE0_05 | >50 | 3.4 | - Metallic<br>- $1.6\times10^{-4}$ | - $R_{4P\_MAX}$: 5.24333 Ω<br>- $R_{4P\_MIN}$: 3.46845 Ω |
|---|---|---|---|---|---|
| 84 | P20119FE0_06 | >50 | 0.322 | - Metallic<br>- $3.01\times10^{-4}$ | - $R_{4P\_MAX}$: 1.36174 Ω<br>- $R_{4P\_MIN}$: 0.32166 Ω |
| 85 | P20119FES_07 | >50 | 0.064 | - Metallic<br>- $1.04\times10^{-3}$ | - $R_{4P\_MAX}$: 0.40296 Ω<br>- $R_{4P\_MIN}$: 0.06204 Ω<br>- Simultaneous signature in $R$ and inductive response<br>- $T_C \sim 262$ K<br>- See Figure S10 and S25 |
| 86 | P20119FE0_08 | >50 | 0.168 | - Metallic<br>- $1.32\times10^{-3}$ | - $R_{4P\_MAX}$: 0.21538 Ω<br>- $R_{4P\_MIN}$: 0.15962 Ω<br>- $T_C \sim 261$ K<br>- Simultaneous signature in $R$ and inductive response<br>- Abrupt jumps in one heating cycle<br>- See Figures S10 and S28 |
| 87 | P20119FE0_09 | >50 | 0.328 | - Metallic<br>- $1.33\times10^{-3}$ | |
| 88 | P20119FE0_10 | >50 | 0.168 | • Metallic<br>• $1.15 \times 10^{-3}$ | |
| 89 | P20119FE0_11 | < 5 | 0.10 | • Metallic<br>• $1.58 \times 10^{-4}$ | • Increase in noise in one channel around $T = 220$ K |
| 90 | P20119FE0_12 | < 5 | 1.31 | • Metallic<br>• $7.75 \times 10^{-4}$ | |
| 91 | P20119FE0_13 | < 5 | 0.196 | • Metallic<br>• $1.37 \times 10^{-3}$ | • Multiple re-entrant transitions<br>• Observation of zero resistance<br>• See Figure S22 |
| 92 | P20119FE0_14 | 5-10 | 0.0918 | • Metallic<br>• $9.66 \times 10^{-4}$ | |
| 93 | P20119FE0_15 | 10-20 | 0.290 | • Metallic<br>• $1.05 \times 10^{-3}$ | • |
| 94 | P20219FE0_16 | 10-20 | 0.587 | • Metallic<br>• $1.37 \times 10^{-3}$ | • Simultaneous signature in $R$ and inductive response |
| 95 | P20219FE0_17 | 10-20 | 0.960 | • Metallic<br>• $1.3 \times 10^{-3}$ | |
| 96 | P20219FE0_18 | 20-30 | 0.090 | • Metallic<br>• $1.85 \times 10^{-3}$ | • Broad, incomplete transition ($\Delta R \sim 1.4\,\%$)<br>• Simultaneous signature in $R$ and inductive response<br>• See Figure S27 |
| 97 | P20219FE0_19 | 20-30 | 0.124 | • Metallic<br>• $1.64 \times 10^{-3}$ | |
| 98 | P20219FE0_20 | 20-30 | 0.191 | • Metallic<br>• $1.84 \times 10^{-3}$ | |
| 99 | P20219FE0_21 | 20-30 | NA | • NA | • No signature in inductive response |
| 100 | P20219FE0_22 | 20-30 | NA | - NA | - No signature in inductive response |
| 101 | P20219FEC_23 | 20-30 | 0.196 | - Metallic<br>- $3.4 \times 10^{-4}$ | |
| 102 | P20219FEC_24 | 20-30 | 0.067 | - Metallic<br>- $1.06 \times 10^{-3}$ | |
| 103 | P20219FEN_25 | 20-30 | 0.884 | - Metallic<br>- $1.21 \times 10^{-3}$ | |
| 104 | P20219FES_26 | 20-30 | 0.066 | - Metallic<br>- $8.61 \times 10^{-4}$ | - Jumps observed during heating |
| 105 | P20219FE0_27 | 20-30 | 0.059 | - Metallic<br>- $6.6 \times 10^{-4}$ | - Minor drop during cooling |
| 106 | P20219FES_28 | 20-30 | 0.074 | <ul><li>Metallic</li><li>$1.27 \times 10^{-3}$</li></ul> | |
| 107 | P20219FE0_29 | 20-30 | 0.885 | <ul><li>Metallic</li><li>$1.17 \times 10^{-3}$</li></ul> | |
| 108 | P20219FE0_30 | 20-30 | 0.630 | <ul><li>Metallic</li><li>$1 \times 10^{-3}$</li></ul> | |
| 109 | P20319FEE_02 | 20-30 | 0.0021 | <ul><li>Metallic</li><li>$8.28 \times 10^{-4}$</li></ul> | |
| | | | **ENCAPSULATED SAMPLES** | | |
| 110 | P20319FEE_03 | 20-30 | 0.108 | <ul><li>Metallic</li><li>$1.39 \times 10^{-4}$</li></ul> | <ul><li>Transition at $T = 65.7$ K</li><li>Observation of zero resistance</li><li>Contact damaged at $T = 60$ K</li><li>See Figure S31</li></ul> |

| 111 | P20319FEE_04 | 20-30 | 1.0 | • Metallic<br>• $4.71 \times 10^{-4}$ | • Incomplete transition at $T = 210$ K |
|---|---|---|---|---|---|
| 112 | P20319FEE_05 | 20-30 | 0.247 | • Metallic<br>• $1.01 \times 10^{-3}$ | • Simultaneous signature in $R$ and inductive response<br>• $T_C = 191.8$ K<br>• See Figure S29 and S31 |
| 113 | P20319FEE_06 | 20-30 | 0.23 | • $7.65 \times 10^{-4}$ | • $T_C = 274$ K<br>• Observation of zero resistance<br>• See Figure S31 and S32 |
| 114 | P20319FEE_07 | 20-30 | 0.288 | • Mixed | • Metal to insulator transition |
| 115 | P20319FEE_08 | 20-30 | 0.38 | • NA | • No signature in inductive response |
| 116 | P20319FEE_09 | 20-30 | 0.320 | • Mixed | • Metal to insulator transition at $T = 210$ K |

| 117 | P20319FEE_10 | 20-30 | 0.046 | • $1.5 \times 10^{-3}$ | • Small abrupt jumps in $R$ |
|---|---|---|---|---|---|
| 118 | P20319FEE_11 | 20-30 | 0.178 | • Metallic<br>• $1.3 \times 10^{-3}$ | • Contacts damaged at $T = 200$ K |
| 119 | P20319FEE_12 | 20-30 | 0.180 | • Metallic<br>• $1.11 \times 10^{-3}$ | |
| 120 | P20319FEE_13 | < 10 | 0.200 | • Metallic<br>• $2.0 \times 10^{-3}$ | |
| 121 | P20319FEE_14 | Not Available | 0.02 | • Insulating | • Contacts damaged at $T = 155$ K |
| 122 | P20319FEE_15 | Not Available | 0.370 | • Metallic<br>• $1.4 \times 10^{-3}$ | |

| 123 | P20319FEE_16 | Not Available | 0.271 | • Metallic | |
| 124 | P20319FEE_17 | Not Available | | • NA | • No signature in inductive response |
| 125 | P20319FEE_18 | Not Available | | • NA | • Signature in inductive response<br>• See Figure S29 |
| 126 | P20519FEE_19 | Not Available | 2.38 | • Metallic<br>• $7 \times 10^{-4}$ | • Signature in inductive response |
| 127 | P20519FEE_20 | Not Available | 0.59 | • $8 \times 10^{-3}$ | • $T_C = 178.8$ K<br>• See Figure 2b in main text |
| 128 | P20519FEE_21 | Not Available | 0.52 | | • Simultaneous signature in $R$ and inductive response<br>• $T_C = 172.5$ K<br>• See Figure 4 in main text |